\documentclass[longbibliography,aps,prx,twocolumn,showpacs,superscriptaddress]{revtex4-1}

\usepackage{graphicx}
\usepackage{dcolumn}
\usepackage{bm}
\usepackage{hyperref}
\usepackage{amsmath}
\usepackage[usenames,dvipsnames]{xcolor}
\usepackage{bm,upgreek}
\usepackage{times}
\usepackage{setspace}
\usepackage{microtype}
\usepackage{datetime}
\usepackage{url}

\newcommand{\bq}{\begin{equation}}
\newcommand{\eq}{\end{equation}}
\newcommand{\bqa}{\begin{eqnarray}}
\newcommand{\eqa}{\end{eqnarray}}

\begin{document}

\title{Order-to-chaos transition in the hardness of random Boolean satisfiability problems}

\author{R\'obert Sumi}\email{equal contribution}
\affiliation{Faculty of Physics, Hungarian Physics Institute, 
Babes-Bolyai University, Cluj-Napoca, Romania}
\author{Melinda Varga}\email{equal contribution}
\affiliation{Department of Physics and the Interdisciplinary 
Center for Network Science and  Applications (iCeNSA),   
University of Notre Dame, Notre Dame, IN, 46556 USA}
\author{Zolt\'an Toroczkai}\email{toro@nd.edu}
\affiliation{Department of Physics and the Interdisciplinary 
Center for Network Science and  Applications (iCeNSA),   
University of Notre Dame, Notre Dame, IN, 46556 USA}
\affiliation{Department of Computer Science and Engineering, 
University of Notre Dame, Notre Dame, IN, 46556 USA}
\author{M\'aria Ercsey-Ravasz} \email{ercsey.ravasz@phys.ubbcluj.ro}
\affiliation{Faculty of Physics, Hungarian Physics Institute, 
Babes-Bolyai University,  Cluj-Napoca, Romania}

\date{\today}

\begin{abstract} 
Transient chaos is an ubiquitous phenomenon characterizing 
the dynamics of phase space trajectories evolving towards a steady state 
attractor in physical systems as diverse 
as fluids, chemical reactions and condensed matter systems. 
Here we show that transient chaos also appears in
the dynamics of certain efficient algorithms searching for solutions of constraint satisfaction 
problems that include scheduling, circuit design, routing, database problems or even Sudoku. 
In particular, we present a study of the emergence of hardness in Boolean 
satisfiability ($k$-SAT), a canonical class of constraint satisfaction problems, by using 
an analog deterministic algorithm based on a system of ordinary differential equations. 
Problem hardness is defined through the escape rate $\kappa$, 
an invariant measure of transient chaos of the dynamical system corresponding to the analog algorithm, 
and it expresses the rate at which the trajectory approaches a solution.
We show that for a given density of constraints and fixed number of Boolean 
variables $N$, the hardness of formulas in random $k$-SAT ensembles has a wide variation, approximable by a lognormal distribution.  
We also show that when increasing the density of constraints $\alpha$, hardness appears
through a second-order phase transition at $\alpha_{\chi}$ in the random 3-SAT ensemble where dynamical
trajectories become transiently chaotic. A similar behavior is found in 4-SAT as well, however, such 
transition does not occur for 2-SAT. This behavior also implies a novel type of transient chaos  
in which the escape rate has an exponential-algebraic
dependence on the critical parameter $\kappa \sim N^{B|\alpha - \alpha_{\chi}|^{1-\gamma}}$ 
with $0< \gamma < 1$. 
We demonstrate that the transition is generated by the appearance of metastable basins in the
solution space as the density of constraints $\alpha$ is increased.  
\end{abstract}


\maketitle

\section{Introduction}

Constraint satisfaction problems arise in many domains of computer 
science, statistical physics, information theory and engineering, or even 
popular puzzles such as Sudoku. In 
these problems there are given $N$ variables, a set of constraints and 
the task is to assign values to the variables such as to satisfy all the 
constraints.  One of the most studied constraint satisfaction problems 
is Boolean satisfiability (SAT) in which all the variables are Boolean 
($x_i\in\{0,1\}$, $i=1,\dots,N$) and the constraints are logical statements 
with a unique truth value, involving these variables. The conjunction 
of these statements is called a formula, and the goal is choosing the
variables such as to satisfy a given formula. According to a fundamental 
theorem of propositional calculus, every such formula $F$ can be 
converted into conjunctive normal form. In CNF a formula is 
expressed as the conjunction (AND, $\wedge$) of $M$ clauses $C_m$, 
$m=1,\ldots,M$, i.e., $F = \bigwedge_{m} C_m$ with each clause 
being a disjunction (OR, $\vee$) of literals (a literal is a variable $x_i$ or
its negation $\neg x_i \equiv \overline{x}_i$) that make up the clause. 
Typical studies focus on $k$-SAT problems in which every clause contains $k$
literals. A simple example with $N=3$, $M=4$ is the set of clauses 
$C_1 = (\overline{x}_1\vee \overline{x}_2 \vee x_3)$, 
$C_2 = (x_1\vee \overline{x}_2 \vee x_3)$, 
$C_3 = (\overline{x}_1\vee x_2 \vee x_3)$ and
$C_4 = (\overline{x}_1\vee \overline{x}_2 \vee \overline{x}_3)$. The solutions 
$(x_1,x_2,x_3)$ that satisfy all four constraints are 
$\{(1,0,1), (0,0,0), (0,0,1),(0,1,1)\}$.

When there are only a small number of constraints, i.e., the 
{\em constraint density} $\alpha = M/N$ is small, the solutions are easily found 
and there are many of them. When $\alpha$ is large, it is easy to show that
the formula cannot be satisfied by any assignment of the variables. However, when
$\alpha$ has intermediate values, finding a solution, or even showing that
one (or none) exists can be very difficult. $k$-SAT for $k \geq 3$ has 
been shown to be NP-hard, i.e., while any candidate solution can be tested to 
satisfy the formula in polynomial time (in $N$) there is no known algorithm 
that would find solutions (or show there aren't any) in polynomial time 
(poly-time). It is also NP-complete \cite{STOC_C71,GareyJohnson79}, which 
means that all problems in the NP complexity class can be reduced in poly-time 
to $k$-SAT (Cook-Levin theorem) and thus the famous $P$ vs $NP$ question can 
be answered from studying $k$-SAT: a poly-time algorithm for $k$-SAT with 
$k \geq 3$ would imply that all problems in NP are tractable ($P=NP$), otherwise 
$P\neq NP$.  For a visual introduction to the $P$ vs. $NP$ 
problem see for example \cite{PvsNP}, and for a
general audience review see \cite{ACM_F09}.
$k$-SAT has major applications in  artificial intelligence, electronic 
design, automation, error-correction, bioinformatics, protein folding, drug-design, etc.  
Although still open, the current strongly held belief is that $P \neq NP$, which means 
that there is no poly-time algorithm to find $k$-SAT solutions for all formulas for $k \geq 3$. 
Since NP-hard problems occur frequently in daily and practical situations, 
there has been a strong impetus to understand the nature of the hardness in such 
problems, as such an understanding may lead to better heuristic algorithms.

In an earlier study \cite{NatPhys_ET11} we opened a novel perspective on SAT 
problems by introducing a new algorithm, as a continuous-time dynamical system 
(CTDS) in form of coupled ordinary differential equations (ODEs), to solve SAT 
problems. This algorithm or analog solver is {\em deterministic}, without backtracks or 
restarts and it is qualitatively different from the other, discrete algorithms. Using this 
solver we were able to demonstrate that constraint satisfaction hardness translates into a
transient chaotic dynamics in the continuous-time dynamical system of the solver.
Note that the appearance of chaos correlating with complexity transitions seems to be
a generic phenomenon, supported by observations also using other, heuristic methods to solve
NP hard problems. Examples include iterated maps \cite{PNAS_ElserRT_07} and the use
of algorithms for finding the ground states of Potts spin-glass type Hamiltonians that 
emulate the NP-hard problem of community detection in networks \cite{PhilMag_HuRN_12, PRE_HuRN_12}. In Ref. \cite{PhilMag_HuRN_12} Hu et al. by map the spin glass Hamiltonian to a dissipative, continuous dynamical system, which was shown to behave chaotically in unsolvable regimes. 
 
As our approach is deterministic and non-heuristic, also makes it possible to bring 
the methods of nonlinear dynamical systems and chaos theory to bear on constraint 
satisfaction problems.
In particular, we have shown that an invariant of the dynamics, 
the so-called {\em escape rate} $\kappa$ is a good measure of hardness of {\em individual} 
SAT instances, which correlated surprisingly well with the subjective hardness ratings
by humans on Sudoku problems \cite{SciRep_ET12}; this measure is the main subject of this paper. Note that  the CTDS 
algorithm is incomplete: it will always find solutions if they exist, but if the instance 
is not satisfiable, then the dynamics will keep running forever. However, the CTDS does solve MAXSAT, i.e., finds solutions which minimize the number of unsatisfied clauses.

The paper is organized as follows. In Section \ref{sec:ana} we present the analog SAT
solver from \cite{NatPhys_ET11} and its properties relevant to the present study. In Section
\ref{sec:tra} we introduce the notion of escape rate as a SAT problem hardness measure 
and show that it goes through a phase transition in the random 3-SAT ensemble as the
constraint density is increased; we also show that this transition is of second order, similar to
the transition in Ising ferromagnets. Section \ref{sec:chaos}  presents an argument for the
nature of this transition while Section \ref{sec:gen} identifies its origin in the appearance 
of certain metastable structures in the solution space.  Section \ref{sec:4sat} shows that the same phenomenon holds for 4-SAT, followed by section \ref{sec:no}, in which we demonstrate that 2-SAT
does not present this chaotic transition. Section \ref{sec:dis} is devoted to conclusions and
discussions.

\section{An analog deterministic SAT solver} \label{sec:ana}

To solve SAT with a continuous-time dynamical system \cite{NatPhys_ET11} we first
reformulate the problem in the space of $N$ continuous variables $s_i$, $i=1,\dots,N$, 
which can take any value in the interval $s_i\in[-1,1]$. Here $s_i = -1 $ is associated 
with the Boolean variable $x_i$ being false ($x_i = 0$) and $s_i =  1$ to it being true 
($x_i =1$). Let 
$\mathbf{C}=\{ c_{mi} \}$ be an $M \times N$ matrix with $c_{mi} = 0$ when the literal
$i$ is missing from clause $C_m$,  $c_{mi}=1$ if $x_i\in C_m$ and  $c_{mi}=-1$ for 
$\overline{x}_i \in C_m$ (clauses containing both a variable and its negation
are automatically discarded as they are always satisfied). The matrix $\mathbf{C}$
fully determines the formula $F$.  One can think of the space $\bm{s}$ as an 
$N$-dimensional hypercube ${\cal H}_N$ with sidelength $2$, with the solutions to the formula $F$ 
lying in the corners of ${\cal H}_N$. The CTDS algorithm to be defined below will, however,
be allowed to visit any point within ${\cal H}_N$ and on its boundary.

To every clause $C_m$ we associate an
analog cost function $K(\bm{s})\in [0,1]$ via $K_m(\bm{s})=2^{-k}\prod_{i=1}^N(1-c_{mi}s_i)$.
For example, for the clause $x_1\vee x_3\vee \overline{x}_9$ we have
$K(\bm{s})=(1-s_1)(1-s_3)(1+s_9)/8$.
Every such product has exactly $k$ non-trivial elements of the form $(1\pm s_i)$, 
corresponding to those literals that are present in $C_m$. Clearly, if $x_i \in C_m$ ($c_{mi} = 1$)
the corresponding term is $(1-s_i)$ and $K_m = 0$ if and only if (`iff') $s_i = 1$ (corresponding to $x_i$ being true).
If $\overline{x}_i \in C_m$ ($c_{mi} = -1$) then the term is $(1+s_i)$ and $K_m = 0$ iff 
$s_i = -1$ (corresponding to $x_i$ being false, or the literal $\overline{x}_i$ being true). Thus
$K_m(\bm{s}) = 0$ in those corners of ${\cal H}_N$, and only those, in which the corresponding
Boolean variables $x_i$ satisfy $C_m$. We are searching for those corner points $\bm{s^*}$  
in which all the $C_m$ clauses are satisfied, i.e., $K_m(\bm{s^*}) = 0$, $\forall m\in\{1,\dots,M\}$,
constituting solutions to the formula $F$. Defining the energy function $V(\bm{s},\bm{a})$
associated with formula $F$ via $V(\bm{s,a})=\sum_{m=1}^Ma_mK_m(\bm{s})^2$, we see 
that as long as $a_m > 0$, $\forall m\in\{1,\dots,M\}$ we have $V(\bm{s^*})=0$ if and only if  
$\bm{s^*}$ is a solution to $F$.  The $a_m > 0$ are auxiliary variables and represent the
weight of a clause in the formula; one can think of them as Lagrange multipliers. 
Thus, we are looking for the lowest/zero energy points of $V$. The search dynamics 
defined in \cite{NatPhys_ET11} is a combination of a gradient descent in $\bm{s}$-space with an 
exponential ascent in the auxiliary $\bm{a}$-space:
\begin{eqnarray}
&&\frac{ds_i}{dt}=-\frac{\partial}{\partial s_i} V(\bm{a},\bm{s}),\;\;\;i=1,\dots,N
\label{sdyn}  \\
&&\frac{da_m}{dt}=a_m K_m(\bm{s}), \;\;\; m=1,\dots,M\label{adyn}
\end{eqnarray}
with positive initial conditions $\bm{a}(0) > 0$ (due to (\ref{adyn}), their positivity is then 
preserved during the dynamics). As shown in \cite{NatPhys_ET11}, this system will always find 
solutions when they exist, as the exponential drive will extract the dynamics from any 
local trap (with the exception of highly symmetric formulas $F$ at low $\alpha$, see the 
Supplementary Information section of \cite{NatPhys_ET11}), and drive the trajectory into one of 
the solutions. Note that all solutions $\bm{s^*}$ are attractive for the dynamics 
\cite{NatPhys_ET11}. In \cite{SciRep_ET12} we show that the dynamics is focused in the sense 
that the least satisfied clause (the one with the largest $K_m$ value) dominates $V$ and the 
dynamics drives exponentially fast the variables towards satisfying that clause, until another 
clause becomes dominant and so on. 
It is important to note that while the scaling of the analog time $t$ to find solutions is polynomial in $N$  \cite{NatPhys_ET11}, in {\em physical device implementations} the $a_m$ variables represent voltages or currents and thus the energetic resources needed to find solutions will become exponential for hard formulas which, is of course necessary, assuming the strongly held belief P$\neq$NP. However, while
one does not know how to generate time, we can generate energy (within limits), and thus systems like
\eqref{sdyn}-\eqref{adyn} become realistic candidates for SAT solving devices (analog circuits).  

As a practical note, during simulations one does not have to wait until the trajectory approaches asymptotically a fixed point solution (the SAT solution). Because of the mapping, the Boolean variable $x_i$ has a unique truth assignment whenever $-1 \leq s_i < 0$
(false) or $0 < s_i \leq 1$ (true), so every point $\mathbf{s} \in {\cal H}_N$ 
of the hypercube (inside and on the boundary) has a corresponding unique Boolean assignment. The SAT solution is in the corner of an orthant (and there is only one corner in an orthant), so there is no point waiting for the trajectory to asymptotically approach that corner, one can just simply check whether the assignment corresponding to that orthant is a solution (recall that checking is fast in NP). 

It is often the case, especially for easy formulas that there 
are several solutions $\bm{s^*}$.  The Hamming distance between two solutions (binary strings) is defined as the number of places in which the corresponding values of the bits differ. If two solutions are unit Hamming distance apart we say that they belong to a {\em solution cluster}.  A solution cluster is therefore formed by solutions that can be connected through single-variable flips always staying within satisfying assignments. 
For example, if  $(s_1^*,s_2^*,s_3^*,...,s_N^*)$ and $(-s_1^*,s_2^*,s_3^*,...,s_N^*)$ are both 
solutions, these are both part of the same cluster and in this case $s_1$ is called  a {\em free 
variable} as its value is inconsequential to the solution, because the formula is satisfied by the 
other variables. For this reason, all points on the corresponding edge of the hypercube 
$(s_1,s_2^*,s_3^*,...,s_N^*)$ with $s_1\in[-1,1]$ have $V=0$ and the dynamics can be attracted 
to any point on this edge (see \cite{NatPhys_ET11}, Supplementary Information Section A about attracting sets). 

Another frequently discussed notion in the literature
is that of a {\em frozen} variable. A variable is said to be frozen in a given cluster, if it
takes on the same value for all the solutions within that cluster; as opposed to the free variable, 
its value is crucial for the existence of that solution cluster. In random $k$-SAT, or other 
ensembles of constraint satisfaction problems (such as $k$-coloring, locked occupation problems, 
etc.) the region of constraint density where the problems are hardest for most algorithms, are characterized by all solution clusters having an extensive fraction of frozen variables. 

\section{A chaotic transition in the escape rate} \label{sec:tra}

For satisfiable formulas the solution is always found but the trajectory of the dynamics can be 
transiently chaotic. The harder the formula the longer these chaotic transients, and thus measures 
of transient chaos can be used to quantify the intrinsic hardness properties of individual formulas 
\cite{NatPhys_ET11,SciRep_ET12}. Transient chaos is a well-known phenomenon from the theory of nonlinear 
dynamical systems \cite{PhysRep_TL08,LaiTel11}, observed to occur in many physical systems, 
such as fluids \cite{PhysRevLett.52.2241,PhysRevLett.77.5055,PhysRevLett.96.094501,hof2006finite},
dielectric cavities  \cite{schwefel2004dramatic,PhysRevA.79.013830}, microwave scattering  \cite{PhysRevLett.65.3072}, electric circuits  \cite{PhysRevLett.73.3528}, mechanical systems  \cite{de2006chaos,PhysRevLett.80.700}, NMR lasers \cite{PhysRevLett.73.529}  and chemical reactions  \cite{scott1991transient,wang_physchem}. Transient chaos is associated with the existence of a {\em non-attracting chaotic set (repeller)} in the phase space. It has recently
been shown \cite{PhysRevLett.111.194101} that in undriven dissipative systems the approach to equilibrium is governed by 
several transient chaotic saddles acting as effective, time-varying chaotic sets.

An important aspect of the CTDS is that it is a hyperbolic dynamical system \cite{NatPhys_ET11},
i.e., starting a number  $N(0)$ of trajectories from different initial conditions from the inside of a domain 
containing the chaotic set, the number $N(t)$ of trajectories that are still found within that domain 
after time $t$ is exponentially decaying in time: $N(t) = N(0)e^{-\kappa t}$. The decay rate, or 
{\em escape rate} $\kappa$ is an invariant characteristic of the chaotic set  \cite{Kadanoff01021984, chaosbook} and it describes the 
average rate at which individual trajectories escape to the attractors
(which might be simple attractors, chaotic attractors, or attractors at infinity). Alternatively, we can say that
the probability of a typical trajectory not escaping to the attractor (in our case a SAT solution) 
until time $t$ decays exponentially \cite{PhysRep_TL08,LaiTel11,PNAS_KT84}: $q(t)\sim e^{-\kappa t}$.
Fig. S9 in the Supplementary Information of Ref \cite{NatPhys_ET11} shows such a decay 
for our CTDS (\ref{sdyn})-(\ref{adyn}). Note that in our case $\kappa$ is a characteristic of the 
dynamical system (\ref{sdyn})-(\ref{adyn}), i.e., of a single formula $F$. One can think of the average
lifetime $\tau \sim \frac{1}{\kappa}$ of chaotic transients as the expected solution time for a 
given solvable formula taken by our algorithm when started from a random initial condition.

It is important to note that in order to estimate the hardness of a formula, one does not need to wait until the trajectories actually find solutions, $\kappa$ can be estimated from the statistics of the trajectories within a domain inside the hypercube ${\cal H}_N$. Note that when measuring $\kappa$ we are launching many trajectories from random points from within this domain. The domain has to be large enough to contain a large fraction of the chaotic repeller set. For example, the domain ${\cal D}={\cal H}_N\cap {\cal S}(r)$, where ${\cal S}(r)$ is the $N$-sphere of radius $r = \sqrt{N-1+(k-1)^2/(k+1)^2}$ (where $k=3$ for $3$-SAT, see Fig. S2 in the Supplementary Information of \cite{NatPhys_ET11}), will have this property. 
It is well known from the dynamical systems literature \cite{chaosbook, LaiTel11} 
that after the initial transient trajectories have left the domain, the remaining trajectories are longer lived and are sampling the chaotic set, before themselves also eventually leaving the domain. These trajectories are chaotic and experience the flow structure around the repeller and contribute to the value of $\kappa$.  Although the chaotic set is an unstable set for solvable SAT problems, its existence is what hinders finding the solutions quickly. Thus, the difficulty or the hardness of the problem can be estimated from the statistics of the trajectories inside a large-enough domain in the hypercube, overlapping with the repeller.

To develop a better understanding of the generic properties of hardness 
as function of $\alpha$, researchers have studied properties 
of statistical ensembles of random formulas with a given $\alpha$. The simplest ensemble 
is the random $k$-SAT ensemble where the formulas are generated uniformly at random. 
This ensemble is the main focus of this paper, but there are also other, frequently studied ensembles such as $k$-XORSAT. In $k$-XORSAT a clause 
sets the sum of $k$ Boolean variables (no negation) modulo 2 to a given value (0 or 1); there are $M$ 
such clauses.  Since $k$-XORSAT
has a special linear structure, it can be solved in poly-time via Gaussian elimination, and thus $k$-XORSAT
is in $P$. We will briefly discuss $3$-XORSAT in the light of the present studies, in Section \ref{sec:dis}. 

The ensemble picture and its similarity to spin-glass models, has attracted 
statistical physics methods  from the theory of strongly disordered systems. In particular, replica 
symmetry and cavity methods  made it possible to obtain a better description
of the structure of the solution space \cite{Science_KS94,PhysRevLett_MZ96,
Science_MPZ02,Nature_Achl2005,Mezard-PRL2005,Mezard-Zecchina-PRL2005,
Krzakala-PNAS,Achlioptas-EPJB,PRE_AZ08,JStatMech_ZM08}. Using these 
statistical ensemble methods, sharp transitions were found in the 
thermodynamic limit ($N,M \to \infty$, $\alpha = const.$) as function of $\alpha$. 
These include the  clustering (or dynamical) transition point $\alpha_{d}$ 
\cite{Mezard-Zecchina-PRL2005,Krzakala-PNAS,Achlioptas-EPJB},   the freezing 
transition $\alpha_f$ \cite{PRE_AZ08,JStatMech_ZM08}, and the SAT/UNSAT satisfiability 
threshold $\alpha_s$ \cite{Science_KS94} etc. It is in the range $\alpha \in [\alpha_{f},\alpha_{s}]$ 
where all known algorithms fail or take exponentially long to solve problems , however, recent numerical results indicate that backtracking survey propagation (BSP) can solve some problems efficiently within a range beyond the freezing transition, for 3-SAT \cite{marino2015backtracking}. 
Beyond $\alpha_s$ the probability for a random SAT formula to be solvable is exponentially small in $N$ and the solvable ones are hard.

In Ref \cite{NatPhys_ET11}  we have shown that with increasing $\alpha$ the trajectories of the CTDS 
become chaotic at some transition density $\alpha_{\chi}$, with the nature of chaos changing 
with $\alpha$ towards, and past the SAT/UNSAT transition point $\alpha_s$. In this paper we present a
more in-depth study of the behavior around $\alpha_{\chi}$ for $k=3$.  
 \begin{figure}[htbp] \begin{center}
\centerline{\includegraphics[width=3.2in]{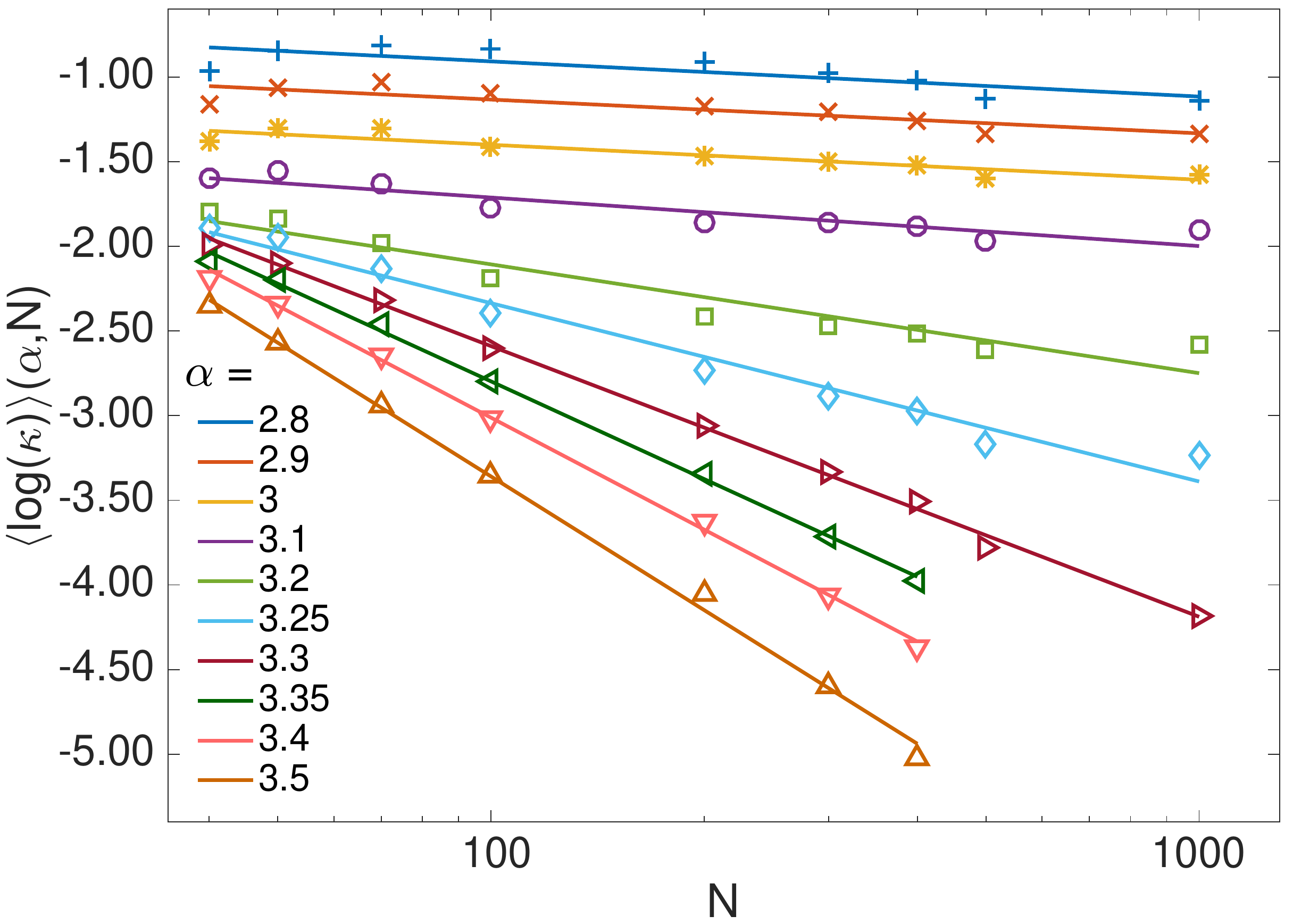}}
\caption{Ensemble averages of the escape rate $\kappa$  calculated on $J$ random instances, for fixed $\alpha$ as function of $N$, in 
logarithmic scale.  The values of $J$ are given in the text.}\label{fig1z} 
\vspace*{-0.5cm} 
\end{center}
 \end{figure}
 
  \begin{figure*}[!htbp] \begin{center}
\includegraphics[width=5.5in]{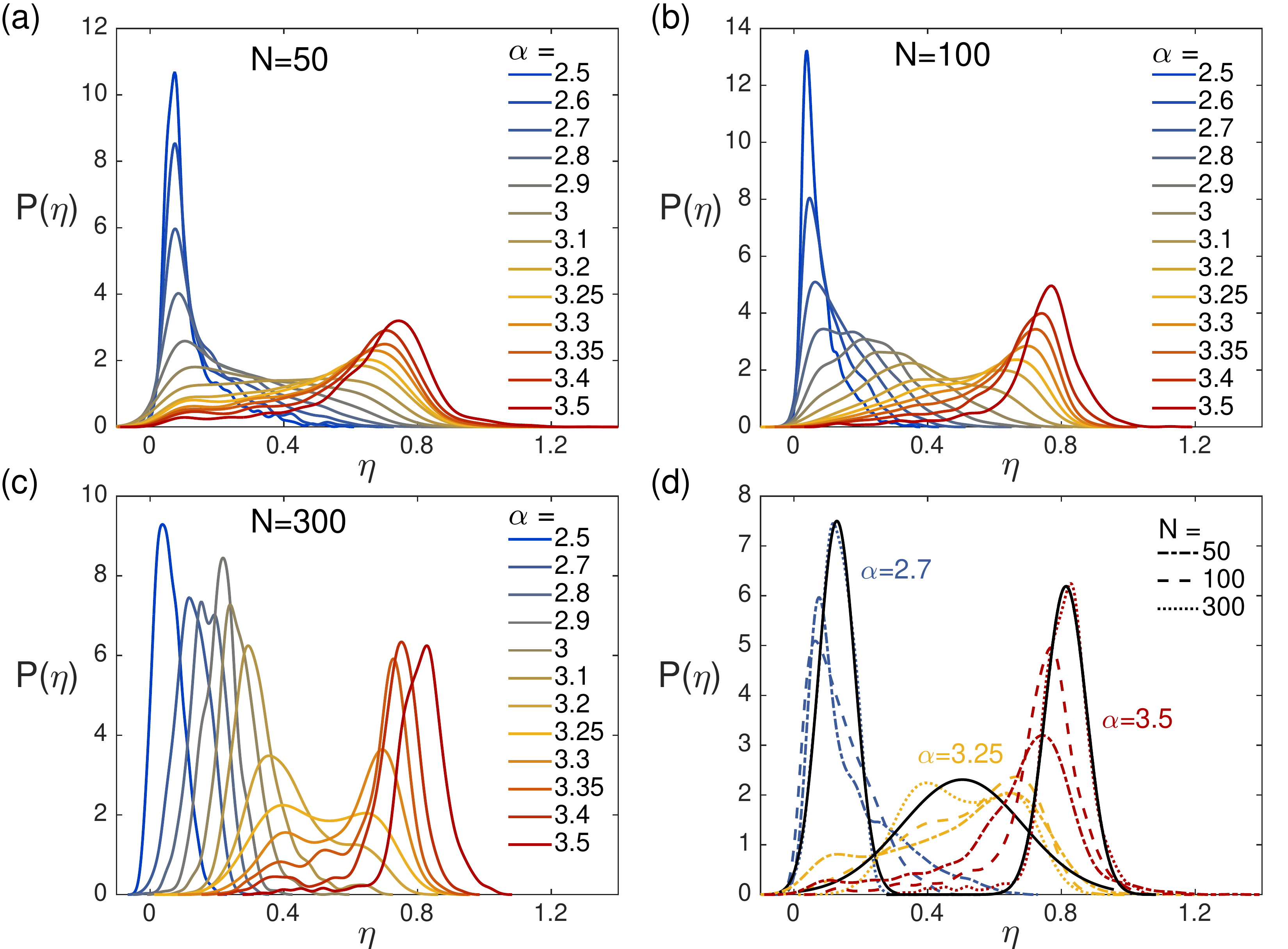}
\caption{Normalized densities $P(\eta)$ of hardness for different $\alpha$ 
values for random satisfiable  $3$-SAT formulas with a) $N=50$, b) $N=100$ and c) $N=300$. 
d) Gaussian fits to the $P(\eta)$ for $N=300$ at $\alpha = 2.7$, $\alpha = 3.25$ and $\alpha = 3.5$, respectively. The distributions became increasingly sharper
with increasing $N$. For a given $N$, when increasing $\alpha$ from smaller values towards larger ones the mean of the distributions shifts to larger values.
  }\label{fig2z} 
\vspace*{-0.5cm} \end{center}
 \end{figure*}
 
Let ${\cal E}(\alpha,N)$ denote the random 3-SAT ensemble of random formulas for fixed $\alpha$ and $N$. Let us denote by $p(t)$ 
the probability that a {\em typical} formula at $(\alpha,N)$ is not solved by time $t$ 
 by our solver.  Numerical evidence indicates the $p(t)$ behaves as $p(t) \sim
e^{- \langle\kappa\rangle t}$, where the decay rate $\langle\kappa\rangle = \langle\kappa\rangle(\alpha,N)$  is the escape rate of a typical formula. It is easy 
to see that $p(t)$ is nothing but the average fraction of formulas from ${\cal E}(\alpha,N)$ 
that have not been solved by time $t$:
the probability that exactly $j$ formulas have not been solved  by time $t$ out of  $J$  independently sampled ones is simply the
binomial distribution $\binom{J}{j} p(t)^{j}[1-p(t)]^{J-j}$. Therefore, the average number of unsolved formulas 
by time $t$ is the average of the binomial, which is $Jp(t)$ and thus the average fraction of unsolved formulas is
simply $p(t)$. In Ref \cite{NatPhys_ET11}  we have shown that this fraction decays exponentially as $e^{-\lambda t}$
which means that $\langle\kappa\rangle = \lambda$. We have also shown that for hard problems 
(e.g., for  $\alpha \geq 4.25$) and large $N$, the exponent $\lambda$ decays polynomially with $N$, i.e., $\lambda \sim N^{-\beta}$, where for example, 
for $\alpha = 4.25$, $\beta \simeq 1.6$.
Fig \ref{fig1z} shows that the power-law decay is true for other constraint densities $\alpha$ 
as well, albeit with an exponent that 
depends on $\alpha$. Fig \ref{fig1z} depicts the logarithm of the escape rate averaged over an ensemble  $\langle \log(\kappa) \rangle(\alpha, N)$ 
as function of $N$, with $\alpha$ fixed, for several $\alpha$ values. In general, for a given $\alpha$ and $N$, $J=1000$ formulas were generated 
from  ${\cal E}(\alpha,N)$ and for each formula we fitted a $\kappa$ value from 
typically $10^3-10^4$ trajectories started from random initial conditions.

As a consequence, the order of magnitude of $\kappa$ for individual formulas  decays polynomially with $N$. 
Even though the ensemble averaged escape rate $\langle \kappa \rangle(\alpha, N)$ is well defined, it might be
that the fluctuations around it are wide, which, as we will show below is actually the case. To better capture individual
fluctuations but remove the leading $N$-dependence from the hardness measure 
we introduce a modified hardness measure, computable for an {\em individual} formula via:
\begin{equation}
\eta=-\frac{\log_{10}\kappa}{\log_{10}N}\;. \label{eta}
\end{equation}

We next use this hardness measure and the finite-size scaling method from statistical mechanics 
(previously applied to SAT in Ref \cite{Science_KS94}) to study the appearance of chaos in the 
behavior of the solver as function of $\alpha$. Note that definition (\ref{eta}) does not eliminate 
completely the $N$-dependence from the hardness measure, only to leading order. 

First, we studied $3$-SAT problems for several values of $N\in[40,1000]$ and $\alpha\in[2.5,4.4]$ 
(see Fig.\ref{fig2z}). We measured  $\eta$ of $J=1000$ (for $N\leq200$), $J=500$ ($N=300,400$) and $J=200$ 
($N=500,1000$) individual random satisfiable formulas. For each formula we ran the dynamics starting 
from $10,000$ (for $N\leq200$), $5,000$ ($N=300,400$) and $2,000$ ($N=500,1000$) random initial 
conditions and estimated the escape rate $\kappa$, and consequently $\eta$. The corresponding distributions 
are shown in Fig. \ref{fig2z}(a-c) for $N=50,100,300$. For small $\alpha$ the $P(\eta)$ distributions are 
sharp and concentrated around their average $\langle \eta \rangle(\alpha,N)$, but with increasing
$\alpha$ the distributions seem to shift to the right, suddenly, becoming concentrated around a larger 
average value. For intermediary $\alpha$ values the distributions are wide and flat, see Fig \ref{fig2z}(d). 
As function of $N$ the distributions become sharper for $\alpha$ away from $\alpha_{\chi} \simeq 3.28$ (where 
the distributions are flat), see Fig \ref{fig2z}(d).

\begin{figure*}[htbp] \begin{center}
\includegraphics[width=5.0in]{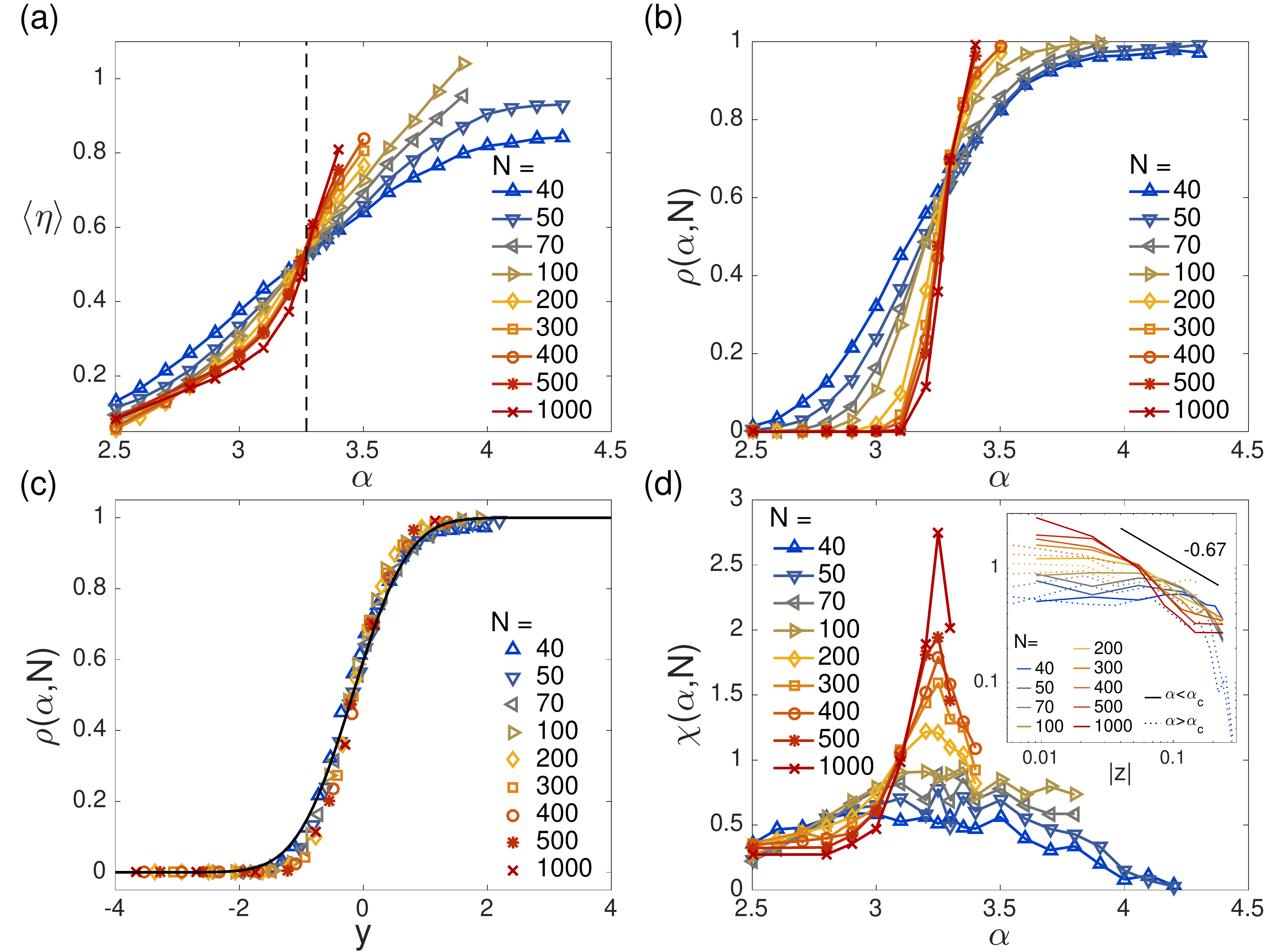}
\caption{(a) Average formula hardness $\langle\eta\rangle(\alpha,N)$ as function of $\alpha$ for different values of $N$. The curves intersect at approximately $\alpha = 3.28$ and $\langle \eta \rangle = 0.5$. b) The fraction of problems $\rho$ with $\eta>\eta_c=0.5$ as function of $\alpha$. 
c) Rescaled $\rho$ curves using Eq (\ref{nu}) with $\nu = 0.5$. The continuous black line is 
the fit $b\,\mbox{erfc}(-y-y_0)$ with $b=0.5$ and $y_{0} = 0.17$ . d) Susceptibility 
(\ref{csu}) as function  of $\alpha$. The inset shows $\chi$ vs. $|z|$. The slope of the fit is -0.67, i.e.,
$\gamma = 0.67$. }\label{fig3z} 
\vspace*{-0.5cm} \end{center}
\end{figure*}

Plotting the average hardness defined by $\langle \eta \rangle(\alpha,N) = \int d\eta\, \eta P(\eta)$ as function of $\alpha$ for each 
$N$ we observe a critical point around $\eta_c\simeq 0.5$ and $\alpha_{\chi} \simeq 3.28$,  where the  
curves intersect each other, shown in Figs. \ref{fig3z}(a). As seen from Fig. \ref{fig2z}(d), 
it is also around this point where the width of the distribution is the largest.  
While for large $N$ it is too costly to do the statistics for 
larger $\alpha$, at small $N$ this was done up to the satisfiability threshold. It shows that the 
hardness constantly increases with $\alpha$ and gets saturated somewhere above $\alpha=4.2$  
near the SAT/UNSAT transition.

 The behavior of the hardness around $\alpha_{\chi}$ has the hallmarks of a second-order phase transition from
 critical phenomena. In analogy with an Ising spin system, if $\alpha$ represents the level of ``drive",
 which is the external magnetic field $H$ in the Ising system, then the average hardness 
 $\langle \eta \rangle(\alpha,N)$ is the system's response, or magnetization per spin $M$ in the Ising model.
 Accordingly 
 \begin{equation}
 \chi (\alpha,N) = \frac{\partial \langle \eta \rangle}{\partial \alpha} \label{csu}
 \end{equation}
 (or $\frac{\partial M}{\partial H}$ in the Ising model) is the corresponding 
``magnetic'' susceptibility. As one can see from Fig \ref{fig3z}(d)  and its inset the susceptibility diverges at $\alpha_{\chi}$
as 
\begin{equation} \label{susc}
\chi \sim |z|^{-\gamma}\;,\;\;\; z\equiv \frac{\alpha - \alpha_{\chi}}{\alpha_{\chi}}\;,\;\;\; \gamma =0.67
\end{equation}
from both sides. 
In Fig. \ref{fig3z}(b) we plot the fraction $\rho$ of formulas with hardness $\eta>\eta_c=0.5$
as function of $\alpha$ for different $N$.
Indeed, one can see that curves for different $N$ intersect each other around the critical value 
$\alpha_{\chi}$ and as $N$ increases the transition between the two phases becomes increasingly sharper. 
Plotting $\rho$ as function of the transformed variable:
\begin{equation}
y=N^{\nu} z \label{nu} 
\end{equation}
the curves can be made to fit on top of each other. The best fit is achieved with $\nu=0.5$. 
\begin{figure*}[htbp] \begin{center}
\includegraphics[width=6in]{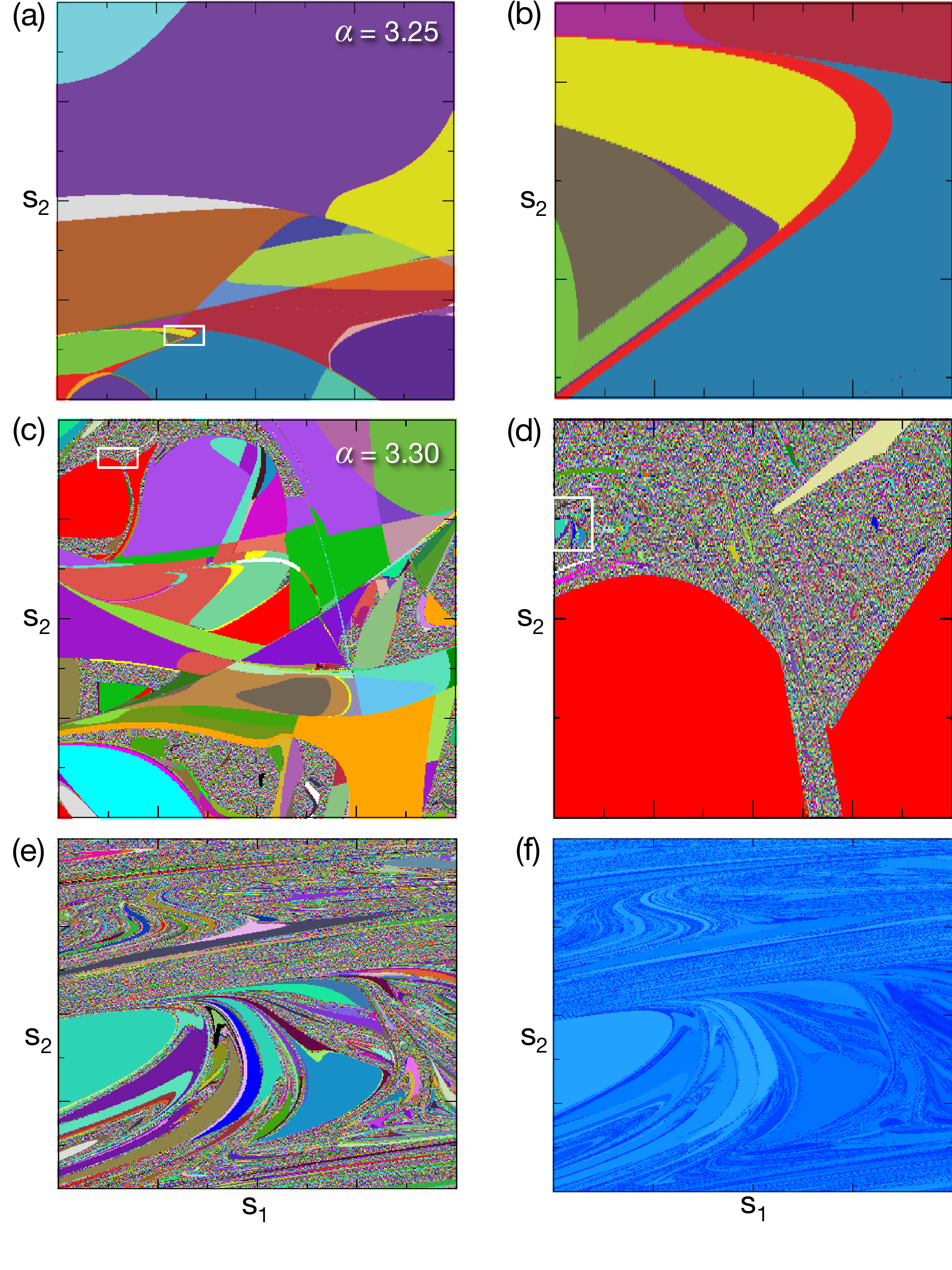}
\caption{Projections of basins of attraction of solutions for a large random formula with $N=1000$ variables. (a) and (b) are at $\alpha=3.25$, 
just before the transition point $\alpha_{\chi}$, whereas (c-f) are at $\alpha = 3.30$, just after $\alpha_{\chi}$. We fixed a random initial condition for all $s_i,\,i\geq3$ and varied only $s_1$ and $s_2$ on a $600\times 600$ grid. The points are colored according to the solution the dynamics flows to.  In (a) and (c) both $s_1$ and $s_2$ span the range $[-1,1]$. (b) is a magnification
of the region in the white rectangle from (a), (d) is the magnification of the region within
the white rectangle from (c) and (e) is a magnification of the white rectangle from (d). (f) shows in shades of blue the time it takes to find the solution for every gridpoint in (e) with darker blue corresponding to longer times. Panels are not to scale. }
\label{fig4z} 
\vspace*{-0.5cm} \end{center}
 \end{figure*}
 
To better interpret the observations above and the corresponding scaling exponents, we will model $P(\eta)$
with an effective, unimodal two-parameter distribution. We also use the observation  
that hardness is bounded from above, i.e., there exists a largest hardness saturation value 
$\eta_{max}$ (in the thermodynamic limit). As discussed earlier, the simulations shown in \cite{NatPhys_ET11} 
generated an exponent of $\beta = 1.6$ for the hardest formulas. which implies that  $\eta \leq \eta_{max} \simeq 1.6$ for all $\alpha$. Thus $P(\eta)$ 
is a distribution with a finite support $\eta \in (0, \eta_{max}]$. A simple choice is that of a truncated Gaussian \cite{url_trun_norm_dist}:
\begin{equation} \label{gauss}
P(\eta) = \frac{A}{\sqrt{2\pi} \sigma}\, e^{-\frac{(\eta-\langle \eta \rangle)^2}{2\sigma^2}}
\end{equation}
where $\langle \eta \rangle = \langle \eta \rangle(\alpha,N)$ is the mean and $\sigma = \sigma(\alpha,N)$
is the width/spread of the distribution. Using this form we can write:
\begin{equation}
\rho(\alpha,N) = \int_{\eta_c}^{\infty} d\eta\,P(\eta) = \frac{A}{2} \mbox{erfc}
\left(\frac{\eta_c - \langle \eta \rangle}{\sqrt{2} \sigma}\right)
\end{equation}
where $\mbox{erfc}(x) = \frac{2}{\sqrt{\pi}} \int_{x}^{\infty} dx\,e^{-x^2}$ is the complementary error function.
The continuous black line in Fig \ref{fig3z}(c) is a fit to a complementary error function 
$\sim\mbox{erfc}(-y-y_0)$. Since the rescaled
$\rho$ curves are well approximated by the erfc near the critical point, we can write:
\begin{equation}
y = N^{\nu} z  =  - \frac{\eta_c - \langle \eta \rangle}{\sqrt{2} \sigma} 
\end{equation} 
or that
\begin{equation} \label{sigma}
\sigma = \frac{|\eta_c - \langle \eta \rangle|}{\sqrt{2} N^{\nu} |z|}
\end{equation}
Eq (\ref{sigma}) shows that the hardness distribution within the ensemble scales as 
$1/\sqrt{N}$ with the number of variables $N$ (the numerator has a bounded variation). In terms of 
$\alpha$ (or $z$), the behavior is more complicated, as for $|z| \to 0$ we have 
$|\eta_c - \langle \eta \rangle| \to 0$ and the functional form depends on the depedence of 
$\langle \eta \rangle(\alpha,N)$ on $\alpha$. However, near the transition point we can write based on 
(\ref{susc}) that 
\begin{equation} \label{cp}
|\eta_c - \langle \eta \rangle| \sim |z|^{1-\gamma},
\end{equation} 
and thus $\sigma \sim |z|^{-\gamma}$, i.e.,
diverges at the critical point, which is consistent with the sudden widening of the distributions seen in Fig \ref{fig2z}.
An interesting consequence of (\ref{sigma}) is that near the critical point ($|z| \ll 1$) we have 
$P(\eta) = P(\alpha,N) = P(z,N) = \frac{A}{\sqrt{2\pi} \sigma} e^{N^{2\nu}z^{2}}$, i.e., $P \sim \sqrt{N}z^{\gamma}e^{N z^2}$.

There are some caveats here. First, the density of the order parameter $\eta$, $P(\eta)$ is not necessarily a
Gaussian, it was only approximated that way. E.g.,  in the 1D Ising model, although the distribution of the 
magnetization is unimodal and Gaussian-like, it is a complicated function (expressed with modified Bessel functions) \cite{antal2004}. A possible alternative for the limiting distribution could even be a
generalized Gumbel-like pdf, which works well, for example, in the 2D XY model in a magnetic field, as shown by Portelli et al. \cite{PhysRevE_PortelliHSB_01}.
Second, because of 
normalization \cite{url_trun_norm_dist},
$A$ in (\ref{gauss}) will also depend on $\langle \eta \rangle$, $\sigma$ and $\eta_{max}$, however, the dependence
is fairly weak. From Fig \ref{fig3z} we find $A \simeq 1.2$ instead of $A=1$ which would be for a non truncated 
Gaussian. Third, the width of the distribution $\sigma$ at the critical point cannot diverge because the whole
distribution is on a bounded support; however, this would be taken care of by the corrections to the Gaussian 
in the true form of $P(\eta)$.

\section{Transient chaos for $\alpha \geq \alpha_{\chi}$} \label{sec:chaos}

While numerical errors are inevitable when estimating the escape rates, the critical value 
$\eta_c=0.5$ has a deeper meaning. 
\begin{figure*}[!htbp] \begin{center}
\includegraphics[width=5.5in]{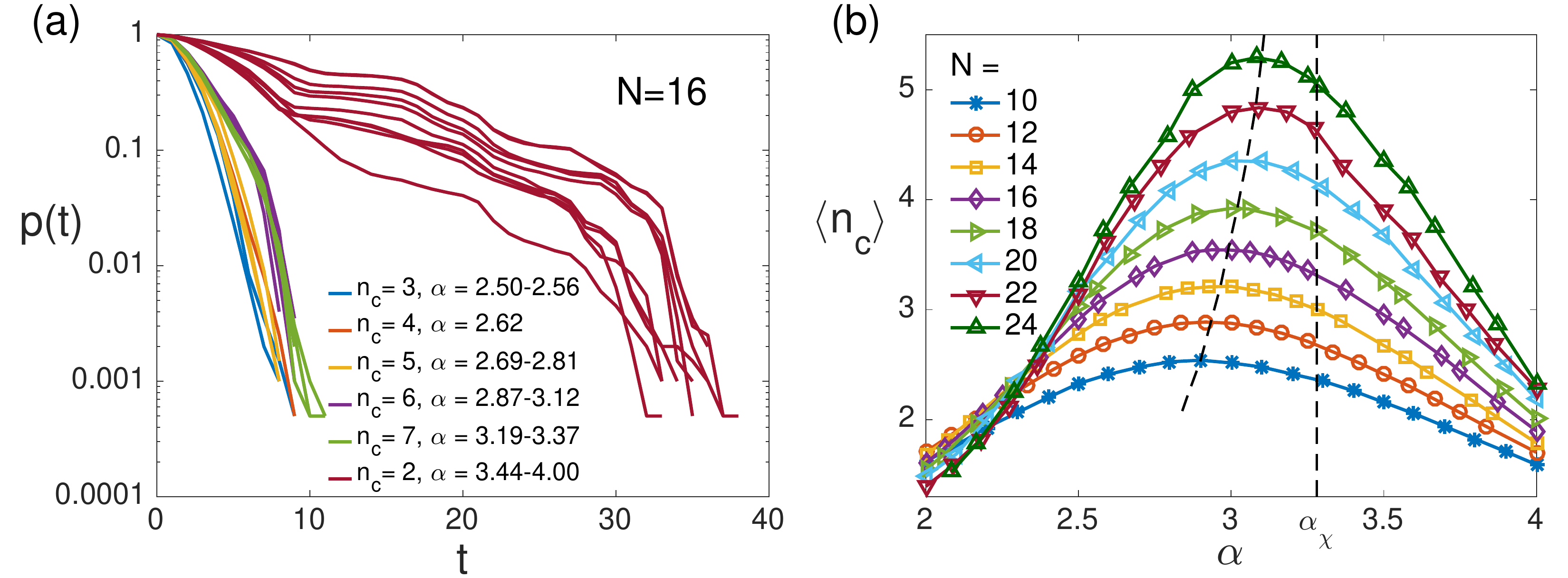}
\caption{a) Distribution of transient times $p(t)$ for a $3$-SAT instance with $N=16$ variables and increasing number of 
constraints. Long chaotic transients appear when the number of solution clusters $n_c$ suddenly decreases (see legend). b) The average number of solution clusters $\langle n_c\rangle$ as function of $\alpha$ measured on $10,000$ 3-SAT instances with different number of variables $N$.  The dashed line connecting the maximum of curves shifts to the right as $N$ increases, slowly converging to $\alpha_{\chi}$.}\label{fig6z} 
\vspace*{-0.5cm} \end{center}
 \end{figure*}
As mentioned above, the search  dynamics moves in the $\bm{s}\in {\cal H}_N$ hypercube and 
solution clusters correspond to attractors, which are most of the time not single points but whole 
subspaces in which every point has potential energy $V=0$. In the absence of chaos  the trajectory 
directly flows into an attractor on a path shorter than the diagonal of the hypercube $2\sqrt{N}$, so 
this length and consequently the characteristic time for finding the solution $\tau$ should scale 
with  an exponent smaller than $0.5$ (the dynamics is accelerated on average
due to (\ref{adyn})). When trajectories become longer 
than the diagonal, the scaling factor of the length and also of the time spent along these trajectories 
is larger than $0.5$, which indicates the appearance of more complicated trajectories. Therefore,  
it is expected that the order-chaos phase transition appears at $\eta_c=0.5$. 
Accordingly, for infinite size formulas (in the thermodynamic limit) the 
probability of a formula generating chaotic dynamics for our CTDS goes to zero for 
$\alpha<\alpha_{\chi}$ and to unity for $\alpha > \alpha_{\chi}$.

If this transition indeed corresponds to the appearance of chaos, then taking any large random  formula 
and gradually increasing $\alpha$ by  adding new  constraints, chaotic dynamics  should  appear close to the threshold. In Figs \ref{fig4z}(a,b) we plot a basin map for a large random instance with $N=1000$ variables at constraint density $\alpha=3.25$. 
Even though the basins have complicated structures, they have piecewise smooth boundaries, revealed upon 
subsequent magnifications.  
Figs \ref{fig4z}(c,d,e) show a similar case but for a random formula obtained  after adding new constraints to the existing ones such as to reach $\alpha=3.30 > \alpha_{\chi}$. 
One can notice the fractal basin boundaries: upon subsequent magnifications the images just show more of the fractal patterns. 
Fig \ref{fig4z}(f) shows the time needed to find a solution from every grid-point. Darker blue corresponds to more time taken by the trajectory. Note that a solution is identified from the orthant which the trajectory enters; as explained in Section \ref{sec:ana}, one can immediately check for
the solution as soon as the trajectory enters a new orthant, since checking is fast,
linear-time cost. In Fig \ref{fig4z} we are coloring initial points by the solutions, not by the clusters, the cluster basin boundary plots have fewer colors, however, they share same behavior as the solution boundary plots, see Fig. 2 in Ref \cite{NatPhys_ET11}.

\section{Metastable energy basins as chaos generators} \label{sec:gen}

Since the attractors of the CTDS correspond to solution clusters, the appearance of chaos has to be 
related to the structure of the solution space. At small $\alpha$ values there are many solutions grouped in only one large solution cluster and possibly a number of small ones. 
Detecting all of  them is computationally costly, it is only possible for small SAT instances 
($N\leq24$). Taking a few formulas with $N=16$ variables and increasing $\alpha$ we monitored how 
the number and size of solution clusters is related to the dynamical 
behavior of the CTDS. We find that having a large number of solution clusters is not a sufficient  condition for chaotic dynamics. 
Instead, it seems that chaos appears when solution clusters start to disappear (Fig.\ref{fig6z}) and metastable (non-solution) energy basins appear. 
A metastable or non-solution energy basin is one from which the trajectory can only escape by increasing its
energy (computed as $E = \sum_{m} K_m^2$), but there is no solution in it (no point with $E = 0$ inside the basin). 
In the energy landscape this would correspond to having deep energy valleys 
without solutions at their bottom; these act as temporary traps from where the dynamics will take time to  escape.
Fig \ref{fig6z}(a) shows the fraction of trajectories $p(t)$ that have not yet found a solution by time $t$, for a formula with
$N=16$, but increasing constraint density. The constraint density was increased by adding random clauses to the existing
ones. For every case (curve) we measured the number of solution clusters $n_c$. One can see that as long as the number
of solution clusters kept increasing the $p(t)$ curves dropped fast. However, as soon as the number of clusters decreased suddenly 
from 7 to 2, long transients appeared and the decay of $p(t)$ has slowed. The following gives a simple scenario by which 
a solution cluster disappears, but remains as a metastable basin. Consider a cluster that has three frozen 
spin variables (see end of Section \ref{sec:ana} for the definition of a frozen variable), for example the first three: $s_1 = 1$, $s_2=1$ and $s_3 = 1$. Adding the new clause 
$\overline{x}_1 \vee \overline{x}_2 \vee \overline{x}_3$, 
it will make all the previous solutions in this cluster non-solutions
for the new formula that includes the added clause. This addition ``lifts'' up uniformly by unity the energy of all the points in the cluster,
thus preserving the energy distribution within the cluster and hence its basin character. When the trajectory gets trapped in this metastable cluster, it will
take time to exit, after some of the auxiliary variables have sufficiently increased.

 We then measured the average number of solution clusters in $10^4$ SAT formulas for $N\in[10,24]$ as function of $\alpha$, with the results shown in Fig  \ref{fig6z}(b).  We can see that the $\alpha'$ where the 
average $\langle \eta_c \rangle$ reaches its maximum is around $3$, but slowly shifts towards larger values with increasing $N$ (see also \cite{PRE_AZ08}). A similar behavior was reported in 
 \cite{thesis_wang} using an image computation method. The presented arguments suggest that $\alpha'$ will converge towards $\alpha_{\chi}$ for large $N$.    

\section{Chaotic transition for 4-SAT} \label{sec:4sat}

\begin{figure*}[!htbp] \begin{center}
\includegraphics[width=5.5in]{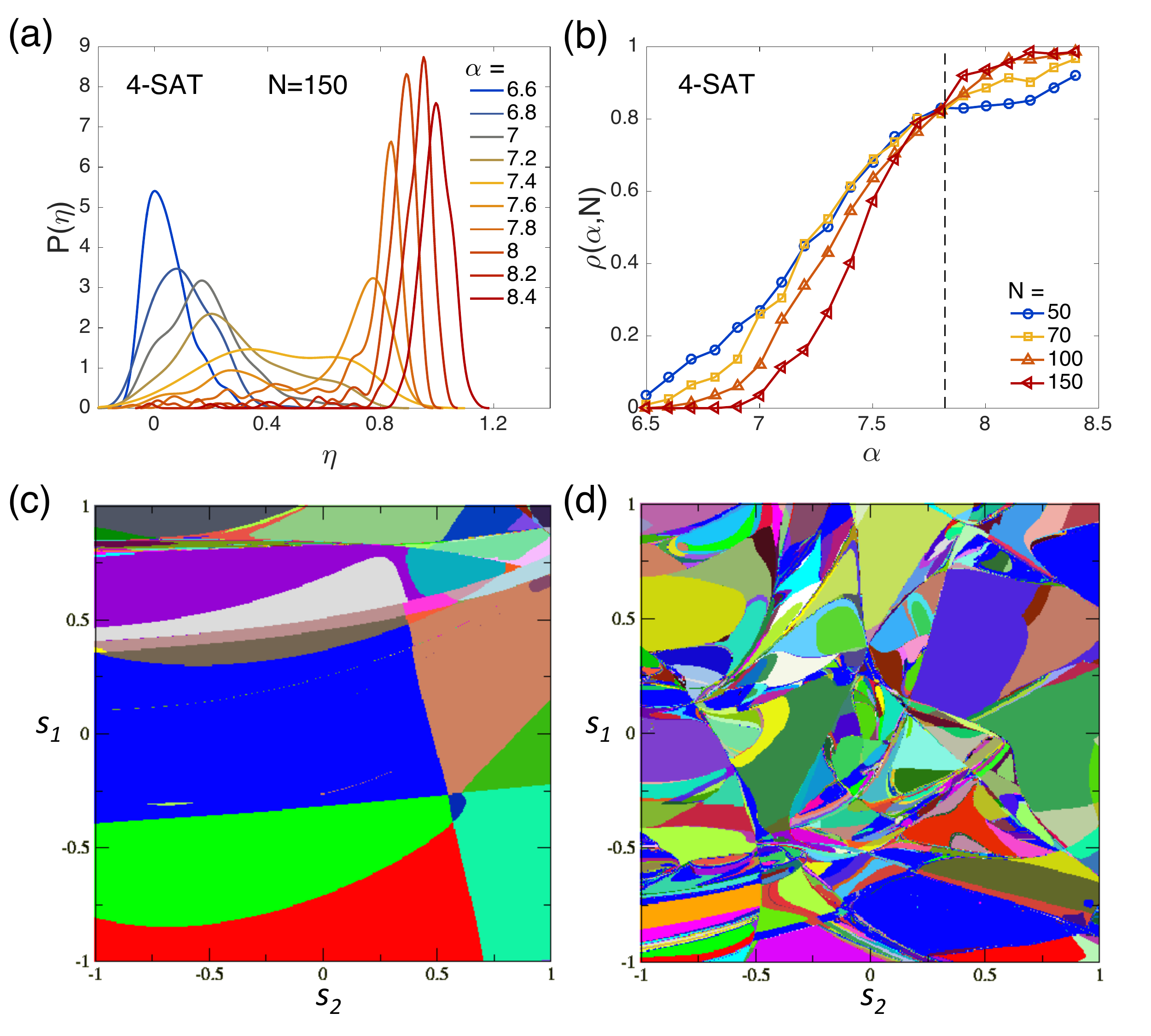}
\caption{a) Normalized densities $P(\eta)$ of hardness for different $\alpha$ values for random satisfiable $4$-SAT instances with $N=150$ variables. b) The fraction of problems $\rho$ with $\eta>\eta_c=0.5$ as function of $\alpha$.  Projections of basins of attraction of solutions for a large random $4$-SAT formula with $N=1000$ variables at c) $\alpha=7.80$ and d) $\alpha=7.90$.  }\label{fig4sat} 
\vspace*{-0.5cm} \end{center}
 \end{figure*}

Although $k$-SAT is NP-complete for $k\geq3$, there are studies arguing that there are qualitative differences between the hardness of $3$-SAT and $4$-SAT   \cite{StatMech_MRTS08} caused by the properties of the solution space. There are heuristic algorithms which efficiently solve some of the $3$-SAT problems even from the frozen region, but they work less well for $k\geq4$  \cite{marino2015backtracking}. For this reason we performed a similar analysis on the random $4$-SAT ensemble. We observe the same type of chaotic transition, in this case around $\alpha_{\chi}\simeq 7.85$, well below the satisfiability threshold  at $\alpha_s=9.931$. Simulations are more costly, so we calculated the escape rate $\kappa$ and hardness $\eta$  for $J=200$ problems of size $N=50, 70, 100, 150$ and $\alpha\in[6.0,8.4]$. The number of initial conditions used in each problem for calculating the  escape rate was $5000$.
In Fig.\ref{fig4sat}(a) we show the distribution $P(\eta)$ at different $\alpha$ values for $N=150$. This figure is very similar to Figs.\ref{fig2z}(a),(b),(c) obtained for $3$-SAT: around a critical $\alpha$ the distributions become wider  with their mean value being again around $\eta_c=0.5$. In Fig.\ref{fig4sat}(b) the curves showing the fraction $\rho$ of problems with hardness larger than $\eta_c=0.5$ indicate a phase transition somewhere in the region $\alpha_{\chi}\in[7.80,7.90]$. Although the simulations are costly and the statistics is not as good as for $3$-SAT, to check if the critical value is indeed in this region, we studied a large random $4$-SAT instance with $N=1000$ variables. Adding more and more constraints, if $N$ is large enough, chaos should appear around the critical value of $\alpha_{\chi}$, as also seen for $3$-SAT in Fig.\ref{fig4z}. Indeed, figures \ref{fig4sat}(c,d) show that the basins of attraction have smooth boundaries at $\alpha=7.80$, but show fractal like features at $7.90$, indicating the appearance of chaos. The 4-SAT example shows that since our algorithm is not dependent on the specificities of the clauses, it can be used to study any SAT problem, even of mixed-type: indeed, 
Sudoku studied in \cite{SciRep_ET12} is a mixed $k$-SAT problem with $k \leq 9$.

\section{No chaotic transition for 2-SAT} \label{sec:no}

\begin{figure*}[!htbp] \begin{center}
\includegraphics[width=5.5in]{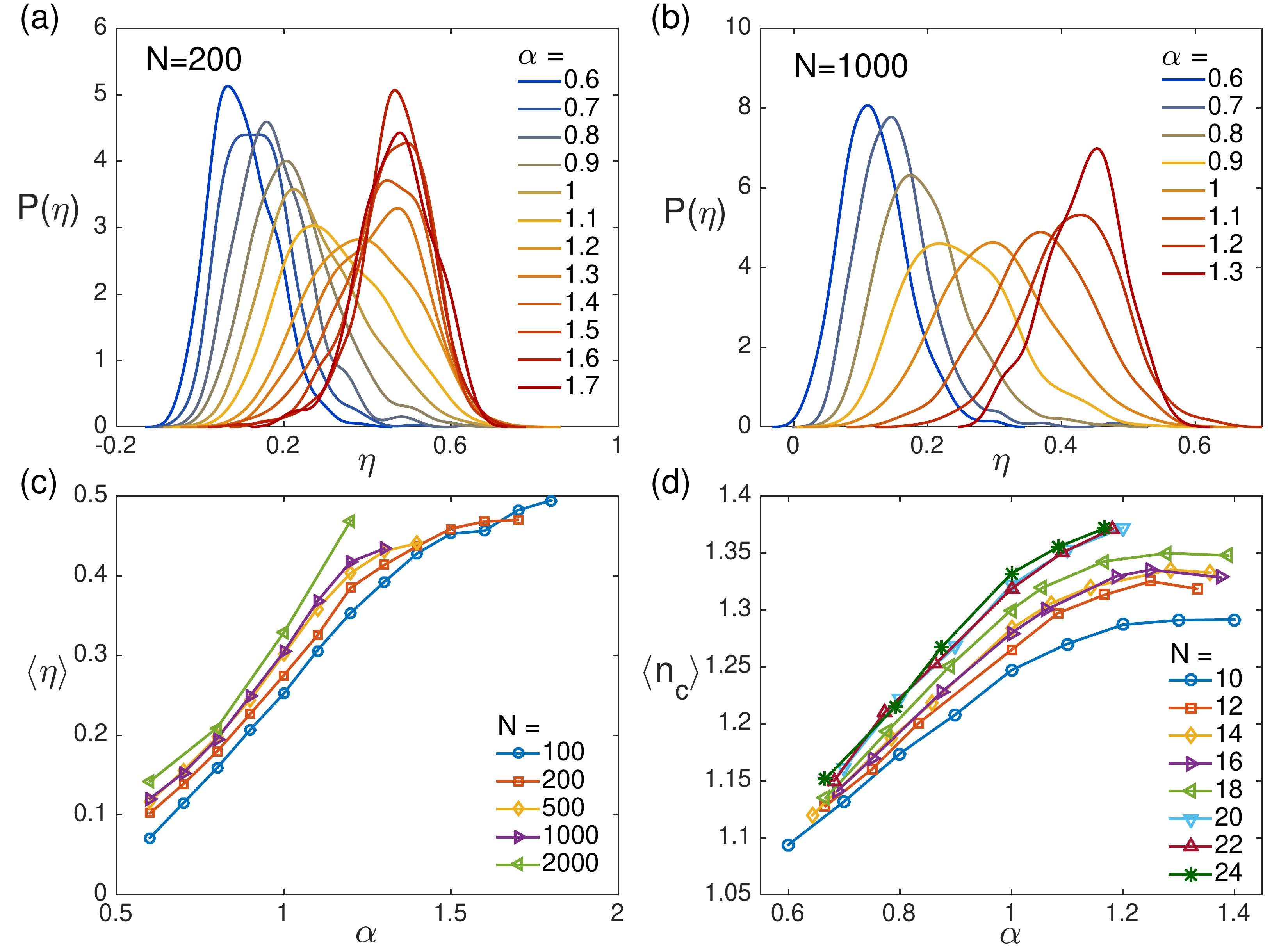}
\caption{Normalized distribution $P(\eta)$ of hardness measures of random satisfiable $2$-SAT instances with different values of $\alpha$ (see legend) and a) $N=200$, b) $N=1000$. c) The average hardness $\langle \eta \rangle>$ and d) average number of solution clusters $\langle n_c\rangle$ as function of $\alpha$ for several values of $N$ (see legends).}\label{fig7z} 
\vspace*{-0.5cm} \end{center}
 \end{figure*}

Next we studied the same statistics, however, for random 2-SAT ensembles. 2-SAT is in $P$ that is, 
there are polynomial-time algorithms that find its solutions or show that there aren't any.   
The distributions of hardness measures for $N=200,1000$ are shown in Fig.\ref{fig7z}(a,b) and 
the average hardness $\langle \eta \rangle$  is plotted as function of $\alpha$ for different values of 
$N$ in Fig.\ref{fig7z}(c). We  see the naturally expected  increase in hardness with increasing $\alpha$, 
however, the transition is missing. The $\langle \eta \rangle$ curves as function of $\alpha$ do not 
intersect each other, the hardness increases with $N$ but rapidly saturates (there is only a very 
small difference between curves for $N=1000$ and 2000) and stay well below the value $\eta_c = 0.5$. Recall that going beyond $\eta_c = 0.5$ signals the fact that the trajectories are much longer
than the typical direct distance between two points in the hypercube ${\cal H}_N$; this does not happen here, statistically. 
However, there can be specific 2-SAT problems {\em for small $N$} for which their hardness goes a little beyond $0.5$ 
as seen from the tails of $P(\eta)$ in Figs \ref{fig7z}(a,b) for large $\alpha$. 
We note that the satisfiability threshold is at $\alpha_s = 1$ in 2-SAT, so for $\alpha$ values within the 
SAT region
only the very tail of the distributions and for smaller $N$ reach barely above $\eta_c = 0.5$.
Just as in the case of 3-SAT, however, with increasing $N$, the $P(\eta)$ distributions become sharper, and
thus the probability that we select randomly a formula with $\eta > 0.5$ within the SAT region goes to zero.
When plotting the number of solution clusters (for satisfiable instances) as function of $\alpha$, as shown in Fig \ref{fig7z}(d) 
we see that for larger $N$ it is only increasing within the SAT (satisfiable) region, then it saturates afterwards. 

Thus chaos disappears in the thermodynamic limit in the SAT region. Of course for unsatisfiable formulas the dynamics will
be permanently chaotic, so the SAT/UNSAT transition point $\alpha_s$ in this case coincides with the chaotic transition point
$\alpha_{\chi}$.

\section{Summary and discussion} \label{sec:dis}
    
We have shown that the escape rate, a dynamic invariant of transient chaos exhibited by the deterministic analog solver provides a suitable 
measure of formula hardness for individual formulas. It is sensitive to changes in the solution space, however, unlike 
hardness measures that are based on solution times, does not require the actual solution of the formula: it is a statistical measure describing the rate of the decay of the number of trajectories that can be estimated from within a domain in the phase space.  

We have provided further numerical evidence that the typical time $\tau \sim \kappa^{-1}$ needed by the analog solver to find a solution for a 
3-SAT formula at a given constraint density $\alpha$ scales polynomially with the number of variables $N$:
\begin{equation}
\tau \sim N^{\eta} \label{anapoly}
\end{equation}
where the exponent $\eta$ depends on both $\alpha$ and $N$, but in a way that it stays bounded, $\eta \in (0,\eta_{max}]$,
where it was found in \cite{NatPhys_ET11} that $\eta_{max}\simeq 1.6$. 
It is important to mention that while the scaling (\ref{anapoly}) indicates that the solver finds solutions in polynomial analog time
(not just on average but also worst case, see \cite{NatPhys_ET11}), this is done at the cost of exponential fluctuations in
energy (due to the exponentially grown auxiliary variables), as shown in  \cite{NatPhys_ET11}, adding further numerical support to the conjecture that $P \neq NP$.   The advantage of this analog approach is in possible non-Turing, special purpose physical implementations 
because  it provides a way to trade search time for energy; while we do not have the ability to generate time, we do know how to generate energy, at least to within reasonable bounds.

We have shown that constraint satisfaction hardness appears in the form of transiently chaotic search trajectories
by the deterministic CTDS.  Transient chaos appears through a second-order phase transition within the random 3-SAT ensemble of
formulas at a critical $\alpha_{\chi} \simeq 3.28$ such that almost no formulas have chaotic trajectories below $\alpha_{\chi}$ and almost all formulas have chaotic
trajectories above $\alpha_{\chi}$, in the thermodynamic limit. 

We have also shown, that surprisingly, for a given finite $N$ and at a fixed $\alpha$ the hardness distribution
is Gaussian-like, which implies that the escape rate and thus the typical search time $\tau$ obeys a lognormal-like
distribution indicating that within random formula ensembles there is a wide range of hardness variability 
(we checked for $\alpha$ over the whole range up to $\alpha_s$), 
somewhat questioning the usefulness of the random ensemble approach for finite problems.  However, (\ref{sigma}) shows that
in the thermodynamic limit these distributions approach delta functions. 

The typical hardness value of a formula around the critical point can be expressed from (\ref{cp}) as
$\eta = 0.5 - B_{l} (\alpha_{\chi} - \alpha)^{1-\gamma}$ for $\alpha < \alpha_{\chi}$ and 
$\eta = 0.5 + B_{r} (\alpha - \alpha_{\chi})^{1-\gamma}$ for $\alpha > \alpha_{\chi}$. This implies that the escape rate 
for fixed $N$ has an exponential-algebraic dependence on the constraint density:
\begin{eqnarray}
\kappa = \left\{ 
\begin{array}{ll}
\frac{1}{\sqrt{N}} e^{(B_{l} \log N )(\alpha_{\chi} - \alpha)^{1-\gamma}  }\;, & \alpha < \alpha_{\chi} \\
\\
\frac{1}{\sqrt{N}} e^{-(B_{r} \log N) (\alpha - \alpha_{\chi})^{1-\gamma} }\;, & \alpha > \alpha_{\chi}
\end{array}
\right.
\end{eqnarray}
which is a form similar to what has been found in the theory of superpersistent chaotic transients, see
Ref \cite{LaiTel11}, pg 298, with the exception 
that for superpersistent transients the exponent $1-\gamma$ is negative and thus $\kappa$ 
approaches zero at the critical point,
whereas here it goes through a non-zero value with a jump in its {\em derivative} at the critical point. 

At this point to clarify some of the relationships 
between problem structure, algorithms and computational complexity. The notion 
of hardness refers to both the algorithm and the problem, the two cannot be separated
in this context. Some problems may appear easy for some algorithms and hard 
for some others. This is why we connected the observed behaviors and the phase transition to the
properties of the {\em solution space}.  Also note that the existence and the properties of
these phase transitions depend on the ensemble of formulas selected.   For example,
random $k$-XORSAT has a different phase diagram than random
$k$-SAT, even though any $k$-XORSAT formula can be brought into a CNF SAT
form (in particular, in the case of $3$-XORSAT we use four CNF 3-SAT clauses to represent one parity check equation). Note that in $k$-XORSAT the constraint density $\gamma$ is measured as the ratio between the number of parity-check equations to the number of variables. The relationship
between variables and constraints can be represented as a hypergraph with nodes representing
the boolean variables and hyperedges representing the parity check equations (constraints) 
connecting the corresponding variables (nodes) present in them.  Using this representation, 
M\'ezard, Ricci-Tershengi and Zecchina have proven \cite{mezard2003two} that  
there is a dynamical transition point at $\gamma_d = 0.8185$ when the statistical weights of the hyperloops (of arbitrary size) in this hypergraph becomes non-zero in the thermodynamic limit.
In the Supplementary Information section of Ref \cite{NatPhys_ET11} we have shown that the corresponding chaotic transition point $\gamma_{\chi}$ for $3$-XORSAT coincides with the dynamical transition point
$\gamma_d$ as it is at this point where chaos appears due to the appearance of small hyperloop motifs in the hypergraph. When (even small) hyperloops are present, mutual coupling appears between our equations (\ref{sdyn}), (\ref{adyn}), and this typically leads to chaotic behavior, see Supplementary Information  Sect G and Fig S4 in \cite{NatPhys_ET11}. A classic example which illustrates that mutual coupling between nonlinear equations induces chaos is the Lorenz system 
\cite{ott2002chaos}. 

It is also important to mention that in spite of an enormous initial excitement, 
the phase transition picture does not directly speak to the true nature of the algorithmic barrier 
$P \neq NP$.  
For example, even though 2-SAT is in $P$, it does have 
a phase transition (SAT/UNSAT transition) at constraint density $\alpha =1$ \cite{chvatal1992mick,goerdt1992threshold}. 
While $k$-XORSAT can be solved in polynomial time by Gaussian elimination (it is 
equivalent to a linear set of equations modulo 2) and thus it is in $P$, it presents 
phase transitions including the clustering transition and the SAT/UNSAT transition.
Additionally, we observe the chaotic phase transition in the XORSAT as well.
Moreover, HornSAT, which is solvable in linear time (thus it is also in $P$) \cite{Beeri_1979,dowling1984linear}, was mathematically proven 
 \cite{moore2005continuous} to also present several phase transitions. 
The phase transition picture appears too coarse in describing the essential nature of the algorithmic 
barrier; one needs tools that can attack this question at the level of single formulas.
The escape rate $\kappa$ or the corresponding hardness measure $\eta$ is a dynamical 
invariant measure of any {\em individual} SAT formula. As the nature of the
dynamics (of the transient chaos) changes significantly from one hard formula to another this measure provides us with a sufficiently sensitive tool to study formula hardness as function of  the matrix $C = \{c_{mi}\}$. It also opens a door to a dynamical systems approach to the $P$ vs $NP$ question.  

Finally, note that system \eqref{sdyn}-\eqref{adyn} is not unique; it is quite possible that there are other forms of dynamical systems that ensure that the trajectory does not get stuck in any non-solution attractors and are perhaps, even better suited for physical device implementations than these equations. 
Eqs \eqref{sdyn}-\eqref{adyn} were introduced to be as transparent as possible, while having the necessary properties required by a deterministic analog  SAT solver. However, since the chaotic transition discussed here is a property of the solution-space, on expects that other deterministic versions of the solver would also experience the same transition.

\section*{Acknowledgments}
 
This project was supported in part by the Romanian CNCS-UEFISCDI, research grant PN-II-RU-TE-3-2011-0121 (MER, RS), the GSCE-30260-2015 "Grant for Supporting Excellent Research" of the Babes-Bolyai University, the UNESCO-L'Oreal National Fellowship "For Women in Science" (MER) and
in part by Grant No. FA9550-12-1-0405 from the U.S. Air Force Office of Scientific Research (AFOSR) and the Defense Advanced Research Projects Agency (DARPA) (ZT). We thank Tam\'as T\'el for  
useful discussions.


\begin{thebibliography}{51}%
\makeatletter
\providecommand \@ifxundefined [1]{%
 \@ifx{#1\undefined}
}%
\providecommand \@ifnum [1]{%
 \ifnum #1\expandafter \@firstoftwo
 \else \expandafter \@secondoftwo
 \fi
}%
\providecommand \@ifx [1]{%
 \ifx #1\expandafter \@firstoftwo
 \else \expandafter \@secondoftwo
 \fi
}%
\providecommand \natexlab [1]{#1}%
\providecommand \enquote  [1]{``#1''}%
\providecommand \bibnamefont  [1]{#1}%
\providecommand \bibfnamefont [1]{#1}%
\providecommand \citenamefont [1]{#1}%
\providecommand \href@noop [0]{\@secondoftwo}%
\providecommand \href [0]{\begingroup \@sanitize@url \@href}%
\providecommand \@href[1]{\@@startlink{#1}\@@href}%
\providecommand \@@href[1]{\endgroup#1\@@endlink}%
\providecommand \@sanitize@url [0]{\catcode `\\12\catcode `\$12\catcode
  `\&12\catcode `\#12\catcode `\^12\catcode `\_12\catcode `\%12\relax}%
\providecommand \@@startlink[1]{}%
\providecommand \@@endlink[0]{}%
\providecommand \url  [0]{\begingroup\@sanitize@url \@url }%
\providecommand \@url [1]{\endgroup\@href {#1}{\urlprefix }}%
\providecommand \urlprefix  [0]{URL }%
\providecommand \Eprint [0]{\href }%
\providecommand \doibase [0]{http://dx.doi.org/}%
\providecommand \selectlanguage [0]{\@gobble}%
\providecommand \bibinfo  [0]{\@secondoftwo}%
\providecommand \bibfield  [0]{\@secondoftwo}%
\providecommand \translation [1]{[#1]}%
\providecommand \BibitemOpen [0]{}%
\providecommand \bibitemStop [0]{}%
\providecommand \bibitemNoStop [0]{.\EOS\space}%
\providecommand \EOS [0]{\spacefactor3000\relax}%
\providecommand \BibitemShut  [1]{\csname bibitem#1\endcsname}%
\let\auto@bib@innerbib\@empty
\bibitem [{\citenamefont {Cook}(1971)}]{STOC_C71}%
  \BibitemOpen
  \bibfield  {author} {\bibinfo {author} {\bibfnamefont {S.A.}\ \bibnamefont
  {Cook}},\ }\bibfield  {title} {\enquote {\bibinfo {title} {The complexity of
  theorem-proving procedures},}\ }in\ \href@noop {} {\emph {\bibinfo
  {booktitle} {Proceedings of the Third Annual ACM Symposium on Theory of
  Computing}}}\ (\bibinfo {organization} {ACM},\ \bibinfo {year} {1971})\ p.\
  \bibinfo {pages} {151}\BibitemShut {NoStop}%
\bibitem [{\citenamefont {Garey}\ and\ \citenamefont
  {Johnson}(1979)}]{GareyJohnson79}%
  \BibitemOpen
  \bibfield  {author} {\bibinfo {author} {\bibfnamefont {M.R.}\ \bibnamefont
  {Garey}}\ and\ \bibinfo {author} {\bibfnamefont {D.S.}\ \bibnamefont
  {Johnson}},\ }\href@noop {} {\emph {\bibinfo {title} {{Computers and
  Intractability: A Guide to the Theory of NP-Completeness (Series of Books in
  the Mathematical Sciences)}}}},\ \bibinfo {edition} {first edition}\ ed.\
  (\bibinfo  {publisher} {W. H. Freeman \& Co Ltd},\ \bibinfo {year}
  {1979})\BibitemShut {NoStop}%
\bibitem [{Pvs()}]{PvsNP}%
  \BibitemOpen
  \href@noop {} {\enquote {\bibinfo {title} {\uppercase{P} vs. \uppercase{NP}
  and the computational complexity zoo},}\ }\bibinfo {howpublished}
  {\url{https://youtu.be/YX40hbAHx3s}},\ \bibinfo {note} {published on Aug 26,
  2014}\BibitemShut {NoStop}%
\bibitem [{\citenamefont {Fortnow}(2009)}]{ACM_F09}%
  \BibitemOpen
  \bibfield  {author} {\bibinfo {author} {\bibfnamefont {L.}~\bibnamefont
  {Fortnow}},\ }\bibfield  {title} {\enquote {\bibinfo {title} {The status of
  the \uppercase{P} versus \uppercase{NP} problem},}\ }\href@noop {} {\bibfield
   {journal} {\bibinfo  {journal} {Commun. ACM}\ }\textbf {\bibinfo {volume}
  {52}},\ \bibinfo {pages} {78} (\bibinfo {year} {2009})}\BibitemShut {NoStop}%
\bibitem [{\citenamefont {Ercsey-Ravasz}\ and\ \citenamefont
  {Toroczkai}({2011})}]{NatPhys_ET11}%
  \BibitemOpen
  \bibfield  {author} {\bibinfo {author} {\bibfnamefont {M.}~\bibnamefont
  {Ercsey-Ravasz}}\ and\ \bibinfo {author} {\bibfnamefont {Z.}~\bibnamefont
  {Toroczkai}},\ }\bibfield  {title} {\enquote {\bibinfo {title} {{Optimization
  hardness as transient chaos in an analog approach to constraint
  satisfaction}},}\ }\href@noop {} {\bibfield  {journal} {\bibinfo  {journal}
  {{Nature Physics}}\ }\textbf {\bibinfo {volume} {{7}}},\ \bibinfo {pages}
  {966} (\bibinfo {year} {{2011}})}\BibitemShut {NoStop}%
\bibitem [{\citenamefont {Elser}\ \emph {et~al.}(2007)\citenamefont {Elser},
  \citenamefont {Rankenburg},\ and\ \citenamefont {P.}}]{PNAS_ElserRT_07}%
  \BibitemOpen
  \bibfield  {author} {\bibinfo {author} {\bibfnamefont {V.}~\bibnamefont
  {Elser}}, \bibinfo {author} {\bibfnamefont {I.}~\bibnamefont {Rankenburg}}, \
  and\ \bibinfo {author} {\bibfnamefont {Thibault}\ \bibnamefont {P.}},\
  }\bibfield  {title} {\enquote {\bibinfo {title} {Searching with iterated
  maps},}\ }\href@noop {} {\bibfield  {journal} {\bibinfo  {journal}
  {Proceedings of the National Academy of Sciences, USA}\ }\textbf {\bibinfo
  {volume} {104}},\ \bibinfo {pages} {418--423} (\bibinfo {year}
  {2007})}\BibitemShut {NoStop}%
\bibitem [{\citenamefont {Hu}\ \emph {et~al.}(2012{\natexlab{a}})\citenamefont
  {Hu}, \citenamefont {Ronhovde},\ and\ \citenamefont
  {Zohar}}]{PhilMag_HuRN_12}%
  \BibitemOpen
  \bibfield  {author} {\bibinfo {author} {\bibfnamefont {DD.}\ \bibnamefont
  {Hu}}, \bibinfo {author} {\bibfnamefont {P.}~\bibnamefont {Ronhovde}}, \ and\
  \bibinfo {author} {\bibfnamefont {N.}~\bibnamefont {Zohar}},\ }\bibfield
  {title} {\enquote {\bibinfo {title} {Phase transitions in random potts
  systems and the community detection problem: spin-glass type and dynamic
  perspectives},}\ }\href@noop {} {\bibfield  {journal} {\bibinfo  {journal}
  {Philosophical Magazine}\ }\textbf {\bibinfo {volume} {92}},\ \bibinfo
  {pages} {406--445} (\bibinfo {year} {2012}{\natexlab{a}})}\BibitemShut
  {NoStop}%
\bibitem [{\citenamefont {Hu}\ \emph {et~al.}(2012{\natexlab{b}})\citenamefont
  {Hu}, \citenamefont {Ronhovde},\ and\ \citenamefont {Zohar}}]{PRE_HuRN_12}%
  \BibitemOpen
  \bibfield  {author} {\bibinfo {author} {\bibfnamefont {DD.}\ \bibnamefont
  {Hu}}, \bibinfo {author} {\bibfnamefont {P.}~\bibnamefont {Ronhovde}}, \ and\
  \bibinfo {author} {\bibfnamefont {N.}~\bibnamefont {Zohar}},\ }\bibfield
  {title} {\enquote {\bibinfo {title} {Stability-to-instability transition in
  the structure of large-scale networks},}\ }\href@noop {} {\bibfield
  {journal} {\bibinfo  {journal} {Physical Review E}\ }\textbf {\bibinfo
  {volume} {86}},\ \bibinfo {pages} {066106} (\bibinfo {year}
  {2012}{\natexlab{b}})}\BibitemShut {NoStop}%
\bibitem [{\citenamefont {Ercsey-Ravasz}\ and\ \citenamefont
  {Toroczkai}({2012})}]{SciRep_ET12}%
  \BibitemOpen
  \bibfield  {author} {\bibinfo {author} {\bibfnamefont {M.}~\bibnamefont
  {Ercsey-Ravasz}}\ and\ \bibinfo {author} {\bibfnamefont {Z.}~\bibnamefont
  {Toroczkai}},\ }\bibfield  {title} {\enquote {\bibinfo {title} {{The chaos
  within Sudoku}},}\ }\href@noop {} {\bibfield  {journal} {\bibinfo  {journal}
  {{Scientific Reports}}\ }\textbf {\bibinfo {volume} {{2}}},\ \bibinfo {pages}
  {725} (\bibinfo {year} {{2012}})}\BibitemShut {NoStop}%
\bibitem [{\citenamefont {T\'el}\ and\ \citenamefont
  {Lai}(2008)}]{PhysRep_TL08}%
  \BibitemOpen
  \bibfield  {author} {\bibinfo {author} {\bibfnamefont {T.}~\bibnamefont
  {T\'el}}\ and\ \bibinfo {author} {\bibfnamefont {Y.-C.}\ \bibnamefont
  {Lai}},\ }\bibfield  {title} {\enquote {\bibinfo {title} {Chaotic transients
  in spatially extended systems},}\ }\href@noop {} {\bibfield  {journal}
  {\bibinfo  {journal} {Physics Reports}\ }\textbf {\bibinfo {volume} {460}},\
  \bibinfo {pages} {245} (\bibinfo {year} {2008})}\BibitemShut {NoStop}%
\bibitem [{\citenamefont {Lai}\ and\ \citenamefont {T\'el}(2011)}]{LaiTel11}%
  \BibitemOpen
  \bibfield  {author} {\bibinfo {author} {\bibfnamefont {Y.-C.}\ \bibnamefont
  {Lai}}\ and\ \bibinfo {author} {\bibfnamefont {T.}~\bibnamefont {T\'el}},\
  }\href@noop {} {\emph {\bibinfo {title} {Transient Chaos: Complex Dynamics on
  Finite-Time Scales}}}\ (\bibinfo  {publisher} {Springer},\ \bibinfo {year}
  {2011})\BibitemShut {NoStop}%
\bibitem [{\citenamefont {Gorman}\ \emph {et~al.}(1984)\citenamefont {Gorman},
  \citenamefont {Widmann},\ and\ \citenamefont
  {Robbins}}]{PhysRevLett.52.2241}%
  \BibitemOpen
  \bibfield  {author} {\bibinfo {author} {\bibfnamefont {M.}~\bibnamefont
  {Gorman}}, \bibinfo {author} {\bibfnamefont {P.~J.}\ \bibnamefont {Widmann}},
  \ and\ \bibinfo {author} {\bibfnamefont {K.~A.}\ \bibnamefont {Robbins}},\
  }\bibfield  {title} {\enquote {\bibinfo {title} {Chaotic flow regimes in a
  convection loop},}\ }\href@noop {} {\bibfield  {journal} {\bibinfo  {journal}
  {Phys. Rev. Lett.}\ }\textbf {\bibinfo {volume} {52}},\ \bibinfo {pages}
  {2241} (\bibinfo {year} {1984})}\BibitemShut {NoStop}%
\bibitem [{\citenamefont {Sommerer}\ \emph {et~al.}(1996)\citenamefont
  {Sommerer}, \citenamefont {Ku},\ and\ \citenamefont
  {Gilreath}}]{PhysRevLett.77.5055}%
  \BibitemOpen
  \bibfield  {author} {\bibinfo {author} {\bibfnamefont {J.C.}\ \bibnamefont
  {Sommerer}}, \bibinfo {author} {\bibfnamefont {H.-C.}\ \bibnamefont {Ku}}, \
  and\ \bibinfo {author} {\bibfnamefont {H.E.}\ \bibnamefont {Gilreath}},\
  }\bibfield  {title} {\enquote {\bibinfo {title} {Experimental evidence for
  chaotic scattering in a fluid wake},}\ }\href@noop {} {\bibfield  {journal}
  {\bibinfo  {journal} {Phys. Rev. Lett.}\ }\textbf {\bibinfo {volume} {77}},\
  \bibinfo {pages} {5055} (\bibinfo {year} {1996})}\BibitemShut {NoStop}%
\bibitem [{\citenamefont {Peixinho}\ and\ \citenamefont
  {Mullin}(2006)}]{PhysRevLett.96.094501}%
  \BibitemOpen
  \bibfield  {author} {\bibinfo {author} {\bibfnamefont {J.}~\bibnamefont
  {Peixinho}}\ and\ \bibinfo {author} {\bibfnamefont {T.}~\bibnamefont
  {Mullin}},\ }\bibfield  {title} {\enquote {\bibinfo {title} {Decay of
  turbulence in pipe flow},}\ }\href@noop {} {\bibfield  {journal} {\bibinfo
  {journal} {Phys. Rev. Lett.}\ }\textbf {\bibinfo {volume} {96}},\ \bibinfo
  {pages} {094501} (\bibinfo {year} {2006})}\BibitemShut {NoStop}%
\bibitem [{\citenamefont {Hof}\ \emph {et~al.}(2006)\citenamefont {Hof},
  \citenamefont {Westerweel}, \citenamefont {Schneider},\ and\ \citenamefont
  {Eckhardt}}]{hof2006finite}%
  \BibitemOpen
  \bibfield  {author} {\bibinfo {author} {\bibfnamefont {B.}~\bibnamefont
  {Hof}}, \bibinfo {author} {\bibfnamefont {J.}~\bibnamefont {Westerweel}},
  \bibinfo {author} {\bibfnamefont {T.M.}\ \bibnamefont {Schneider}}, \ and\
  \bibinfo {author} {\bibfnamefont {B.}~\bibnamefont {Eckhardt}},\ }\bibfield
  {title} {\enquote {\bibinfo {title} {Finite lifetime of turbulence in shear
  flows},}\ }\href@noop {} {\bibfield  {journal} {\bibinfo  {journal} {Nature}\
  }\textbf {\bibinfo {volume} {443}},\ \bibinfo {pages} {59} (\bibinfo {year}
  {2006})}\BibitemShut {NoStop}%
\bibitem [{\citenamefont {Schwefel}\ \emph {et~al.}(2004)\citenamefont
  {Schwefel}, \citenamefont {Rex}, \citenamefont {Tureci}, \citenamefont
  {Chang}, \citenamefont {Stone}, \citenamefont {Ben-Messaoud},\ and\
  \citenamefont {Zyss}}]{schwefel2004dramatic}%
  \BibitemOpen
  \bibfield  {author} {\bibinfo {author} {\bibfnamefont {H.G.L.}\ \bibnamefont
  {Schwefel}}, \bibinfo {author} {\bibfnamefont {N.B.}\ \bibnamefont {Rex}},
  \bibinfo {author} {\bibfnamefont {H.E.}\ \bibnamefont {Tureci}}, \bibinfo
  {author} {\bibfnamefont {R.K.}\ \bibnamefont {Chang}}, \bibinfo {author}
  {\bibfnamefont {A.D.}\ \bibnamefont {Stone}}, \bibinfo {author}
  {\bibfnamefont {T.}~\bibnamefont {Ben-Messaoud}}, \ and\ \bibinfo {author}
  {\bibfnamefont {J.}~\bibnamefont {Zyss}},\ }\bibfield  {title} {\enquote
  {\bibinfo {title} {Dramatic shape sensitivity of directional emission
  patterns from similarly deformed cylindrical polymer lasers},}\ }\href@noop
  {} {\bibfield  {journal} {\bibinfo  {journal} {J. Opt. Soc. Am. B}\ }\textbf
  {\bibinfo {volume} {21}},\ \bibinfo {pages} {923} (\bibinfo {year}
  {2004})}\BibitemShut {NoStop}%
\bibitem [{\citenamefont {Altmann}(2009)}]{PhysRevA.79.013830}%
  \BibitemOpen
  \bibfield  {author} {\bibinfo {author} {\bibfnamefont {E.G.}\ \bibnamefont
  {Altmann}},\ }\bibfield  {title} {\enquote {\bibinfo {title} {Emission from
  dielectric cavities in terms of invariant sets of the chaotic ray
  dynamics},}\ }\href@noop {} {\bibfield  {journal} {\bibinfo  {journal} {Phys.
  Rev. A}\ }\textbf {\bibinfo {volume} {79}},\ \bibinfo {pages} {013830}
  (\bibinfo {year} {2009})}\BibitemShut {NoStop}%
\bibitem [{\citenamefont {Doron}\ \emph {et~al.}(1990)\citenamefont {Doron},
  \citenamefont {Smilansky},\ and\ \citenamefont
  {Frenkel}}]{PhysRevLett.65.3072}%
  \BibitemOpen
  \bibfield  {author} {\bibinfo {author} {\bibfnamefont {E.}~\bibnamefont
  {Doron}}, \bibinfo {author} {\bibfnamefont {U.}~\bibnamefont {Smilansky}}, \
  and\ \bibinfo {author} {\bibfnamefont {A.}~\bibnamefont {Frenkel}},\
  }\bibfield  {title} {\enquote {\bibinfo {title} {Experimental demonstration
  of chaotic scattering of microwaves},}\ }\href@noop {} {\bibfield  {journal}
  {\bibinfo  {journal} {Phys. Rev. Lett.}\ }\textbf {\bibinfo {volume} {65}},\
  \bibinfo {pages} {3072} (\bibinfo {year} {1990})}\BibitemShut {NoStop}%
\bibitem [{\citenamefont {Heagy}\ \emph {et~al.}(1994)\citenamefont {Heagy},
  \citenamefont {Carroll},\ and\ \citenamefont {Pecora}}]{PhysRevLett.73.3528}%
  \BibitemOpen
  \bibfield  {author} {\bibinfo {author} {\bibfnamefont {J.F.}\ \bibnamefont
  {Heagy}}, \bibinfo {author} {\bibfnamefont {T.L.}\ \bibnamefont {Carroll}}, \
  and\ \bibinfo {author} {\bibfnamefont {L.M.}\ \bibnamefont {Pecora}},\
  }\bibfield  {title} {\enquote {\bibinfo {title} {Experimental and numerical
  evidence for riddled basins in coupled chaotic systems},}\ }\href@noop {}
  {\bibfield  {journal} {\bibinfo  {journal} {Phys. Rev. Lett.}\ }\textbf
  {\bibinfo {volume} {73}},\ \bibinfo {pages} {3528} (\bibinfo {year}
  {1994})}\BibitemShut {NoStop}%
\bibitem [{\citenamefont {de~Paula}\ \emph {et~al.}(2006)\citenamefont
  {de~Paula}, \citenamefont {Savi},\ and\ \citenamefont
  {Pereira-Pinto}}]{de2006chaos}%
  \BibitemOpen
  \bibfield  {author} {\bibinfo {author} {\bibfnamefont {A.S.}\ \bibnamefont
  {de~Paula}}, \bibinfo {author} {\bibfnamefont {M.A.}\ \bibnamefont {Savi}}, \
  and\ \bibinfo {author} {\bibfnamefont {F.H.I.}\ \bibnamefont
  {Pereira-Pinto}},\ }\bibfield  {title} {\enquote {\bibinfo {title} {Chaos and
  transient chaos in an experimental nonlinear pendulum},}\ }\href@noop {}
  {\bibfield  {journal} {\bibinfo  {journal} {J. Sound and Vibration}\ }\textbf
  {\bibinfo {volume} {294}},\ \bibinfo {pages} {585} (\bibinfo {year}
  {2006})}\BibitemShut {NoStop}%
\bibitem [{\citenamefont {In}\ \emph {et~al.}(1998)\citenamefont {In},
  \citenamefont {Spano},\ and\ \citenamefont {Ding}}]{PhysRevLett.80.700}%
  \BibitemOpen
  \bibfield  {author} {\bibinfo {author} {\bibfnamefont {V.}~\bibnamefont
  {In}}, \bibinfo {author} {\bibfnamefont {M.L.}\ \bibnamefont {Spano}}, \ and\
  \bibinfo {author} {\bibfnamefont {M.}~\bibnamefont {Ding}},\ }\bibfield
  {title} {\enquote {\bibinfo {title} {Maintaining chaos in high dimensions},}\
  }\href@noop {} {\bibfield  {journal} {\bibinfo  {journal} {Phys. Rev. Lett.}\
  }\textbf {\bibinfo {volume} {80}},\ \bibinfo {pages} {700} (\bibinfo {year}
  {1998})}\BibitemShut {NoStop}%
\bibitem [{\citenamefont {J\'anosi}\ \emph {et~al.}(1994)\citenamefont
  {J\'anosi}, \citenamefont {Flepp},\ and\ \citenamefont
  {T\'el}}]{PhysRevLett.73.529}%
  \BibitemOpen
  \bibfield  {author} {\bibinfo {author} {\bibfnamefont {I.M.}\ \bibnamefont
  {J\'anosi}}, \bibinfo {author} {\bibfnamefont {L.}~\bibnamefont {Flepp}}, \
  and\ \bibinfo {author} {\bibfnamefont {T.}~\bibnamefont {T\'el}},\ }\bibfield
   {title} {\enquote {\bibinfo {title} {Exploring transient chaos in an
  \uppercase{NMR}-laser experiment},}\ }\href@noop {} {\bibfield  {journal}
  {\bibinfo  {journal} {Phys. Rev. Lett.}\ }\textbf {\bibinfo {volume} {73}},\
  \bibinfo {pages} {529} (\bibinfo {year} {1994})}\BibitemShut {NoStop}%
\bibitem [{\citenamefont {Scott}\ \emph {et~al.}(1991)\citenamefont {Scott},
  \citenamefont {Peng}, \citenamefont {Tomlin},\ and\ \citenamefont
  {Showalter}}]{scott1991transient}%
  \BibitemOpen
  \bibfield  {author} {\bibinfo {author} {\bibfnamefont {S.K.}\ \bibnamefont
  {Scott}}, \bibinfo {author} {\bibfnamefont {B.}~\bibnamefont {Peng}},
  \bibinfo {author} {\bibfnamefont {A.S.}\ \bibnamefont {Tomlin}}, \ and\
  \bibinfo {author} {\bibfnamefont {K.}~\bibnamefont {Showalter}},\ }\bibfield
  {title} {\enquote {\bibinfo {title} {Transient chaos in a closed chemical
  system},}\ }\href@noop {} {\bibfield  {journal} {\bibinfo  {journal} {J.
  Chem. Phys.}\ }\textbf {\bibinfo {volume} {94}},\ \bibinfo {pages} {1134}
  (\bibinfo {year} {1991})}\BibitemShut {NoStop}%
\bibitem [{\citenamefont {J.Wang}\ \emph {et~al.}(1994)\citenamefont {J.Wang},
  \citenamefont {Soerensen},\ and\ \citenamefont {Hynne}}]{wang_physchem}%
  \BibitemOpen
  \bibfield  {author} {\bibinfo {author} {\bibnamefont {J.Wang}}, \bibinfo
  {author} {\bibfnamefont {P.~G.}\ \bibnamefont {Soerensen}}, \ and\ \bibinfo
  {author} {\bibfnamefont {F.}~\bibnamefont {Hynne}},\ }\bibfield  {title}
  {\enquote {\bibinfo {title} {Transient period doublings, torus oscillations,
  and chaos in a closed chemical system},}\ }\href@noop {} {\bibfield
  {journal} {\bibinfo  {journal} {J. Phys. Chem.}\ }\textbf {\bibinfo {volume}
  {98}},\ \bibinfo {pages} {725} (\bibinfo {year} {1994})}\BibitemShut
  {NoStop}%
\bibitem [{\citenamefont {Motter}\ \emph {et~al.}(2013)\citenamefont {Motter},
  \citenamefont {Gruiz}, \citenamefont {K\'arolyi},\ and\ \citenamefont
  {T\'el}}]{PhysRevLett.111.194101}%
  \BibitemOpen
  \bibfield  {author} {\bibinfo {author} {\bibfnamefont {A.E.}\ \bibnamefont
  {Motter}}, \bibinfo {author} {\bibfnamefont {M.}~\bibnamefont {Gruiz}},
  \bibinfo {author} {\bibfnamefont {Gy.}\ \bibnamefont {K\'arolyi}}, \ and\
  \bibinfo {author} {\bibfnamefont {T.}~\bibnamefont {T\'el}},\ }\bibfield
  {title} {\enquote {\bibinfo {title} {Doubly transient chaos: Generic form of
  chaos in autonomous dissipative systems},}\ }\href@noop {} {\bibfield
  {journal} {\bibinfo  {journal} {Phys. Rev. Lett.}\ }\textbf {\bibinfo
  {volume} {111}},\ \bibinfo {pages} {194101} (\bibinfo {year}
  {2013})}\BibitemShut {NoStop}%
\bibitem [{\citenamefont {Kadanoff}\ and\ \citenamefont
  {Tang}(1984{\natexlab{a}})}]{Kadanoff01021984}%
  \BibitemOpen
  \bibfield  {author} {\bibinfo {author} {\bibfnamefont {L.~P.}\ \bibnamefont
  {Kadanoff}}\ and\ \bibinfo {author} {\bibfnamefont {C.}~\bibnamefont
  {Tang}},\ }\bibfield  {title} {\enquote {\bibinfo {title} {Escape from
  strange repellers},}\ }\href@noop {} {\bibfield  {journal} {\bibinfo
  {journal} {PNAS}\ }\textbf {\bibinfo {volume} {81}},\ \bibinfo {pages} {1276}
  (\bibinfo {year} {1984}{\natexlab{a}})}\BibitemShut {NoStop}%
\bibitem [{\citenamefont {Cvitanovi\'c}\ \emph {et~al.}(2009)\citenamefont
  {Cvitanovi\'c}, \citenamefont {Artuso}, \citenamefont {Mainieri},
  \citenamefont {Tanner},\ and\ \citenamefont {Vattay}}]{chaosbook}%
  \BibitemOpen
  \bibfield  {author} {\bibinfo {author} {\bibfnamefont {P.}~\bibnamefont
  {Cvitanovi\'c}}, \bibinfo {author} {\bibfnamefont {R.}~\bibnamefont
  {Artuso}}, \bibinfo {author} {\bibfnamefont {R.}~\bibnamefont {Mainieri}},
  \bibinfo {author} {\bibfnamefont {G.}~\bibnamefont {Tanner}}, \ and\ \bibinfo
  {author} {\bibfnamefont {G.}~\bibnamefont {Vattay}},\ }\href
  {http://chaosbook.org} {\emph {\bibinfo {title} {Chaos: Classical and
  Quantum}}}\ (\bibinfo  {publisher} {Niels Bohr Institute, Copenhagen},\
  \bibinfo {year} {2009})\BibitemShut {NoStop}%
\bibitem [{\citenamefont {Kadanoff}\ and\ \citenamefont
  {Tang}(1984{\natexlab{b}})}]{PNAS_KT84}%
  \BibitemOpen
  \bibfield  {author} {\bibinfo {author} {\bibfnamefont {L.P.}\ \bibnamefont
  {Kadanoff}}\ and\ \bibinfo {author} {\bibfnamefont {C.}~\bibnamefont
  {Tang}},\ }\bibfield  {title} {\enquote {\bibinfo {title} {{Escape rate from
  strange repellers}},}\ }\href@noop {} {\bibfield  {journal} {\bibinfo
  {journal} {PNAS}\ }\textbf {\bibinfo {volume} {81}},\ \bibinfo {pages} {1276}
  (\bibinfo {year} {1984}{\natexlab{b}})}\BibitemShut {NoStop}%
\bibitem [{\citenamefont {Kirkpatrick}\ and\ \citenamefont
  {Selman}({1994})}]{Science_KS94}%
  \BibitemOpen
  \bibfield  {author} {\bibinfo {author} {\bibfnamefont {S.}~\bibnamefont
  {Kirkpatrick}}\ and\ \bibinfo {author} {\bibfnamefont {B.}~\bibnamefont
  {Selman}},\ }\bibfield  {title} {\enquote {\bibinfo {title}
  {{Critical-behavior in the satisfiability of random boolean expressions}},}\
  }\href@noop {} {\bibfield  {journal} {\bibinfo  {journal} {{Science}}\
  }\textbf {\bibinfo {volume} {{264}}},\ \bibinfo {pages} {1297} (\bibinfo
  {year} {{1994}})}\BibitemShut {NoStop}%
\bibitem [{\citenamefont {Monasson}\ and\ \citenamefont
  {Zecchina}({1996})}]{PhysRevLett_MZ96}%
  \BibitemOpen
  \bibfield  {author} {\bibinfo {author} {\bibfnamefont {R.}~\bibnamefont
  {Monasson}}\ and\ \bibinfo {author} {\bibfnamefont {R.}~\bibnamefont
  {Zecchina}},\ }\bibfield  {title} {\enquote {\bibinfo {title} {{Entropy of
  the k-satisfiability problem}},}\ }\href@noop {} {\bibfield  {journal}
  {\bibinfo  {journal} {Phys. Rev. Lett.}\ }\textbf {\bibinfo {volume}
  {{76}}},\ \bibinfo {pages} {3881} (\bibinfo {year} {{1996}})}\BibitemShut
  {NoStop}%
\bibitem [{\citenamefont {Mezard}\ \emph {et~al.}({2002})\citenamefont
  {Mezard}, \citenamefont {Parisi},\ and\ \citenamefont
  {Zecchina}}]{Science_MPZ02}%
  \BibitemOpen
  \bibfield  {author} {\bibinfo {author} {\bibfnamefont {M.}~\bibnamefont
  {Mezard}}, \bibinfo {author} {\bibfnamefont {G.}~\bibnamefont {Parisi}}, \
  and\ \bibinfo {author} {\bibfnamefont {R.}~\bibnamefont {Zecchina}},\
  }\bibfield  {title} {\enquote {\bibinfo {title} {{Analytic and algorithmic
  solution of random satisfiability problems}},}\ }\href@noop {} {\bibfield
  {journal} {\bibinfo  {journal} {{Science}}\ }\textbf {\bibinfo {volume}
  {{297}}},\ \bibinfo {pages} {812} (\bibinfo {year} {{2002}})}\BibitemShut
  {NoStop}%
\bibitem [{\citenamefont {Achlioptas}\ \emph {et~al.}(2005)\citenamefont
  {Achlioptas}, \citenamefont {Naor},\ and\ \citenamefont
  {Peres}}]{Nature_Achl2005}%
  \BibitemOpen
  \bibfield  {author} {\bibinfo {author} {\bibfnamefont {D.}~\bibnamefont
  {Achlioptas}}, \bibinfo {author} {\bibfnamefont {A.}~\bibnamefont {Naor}}, \
  and\ \bibinfo {author} {\bibfnamefont {Y.}~\bibnamefont {Peres}},\ }\bibfield
   {title} {\enquote {\bibinfo {title} {Rigorous location of phase transitions
  in hard optimization problems},}\ }\href@noop {} {\bibfield  {journal}
  {\bibinfo  {journal} {Nature}\ }\textbf {\bibinfo {volume} {435}},\ \bibinfo
  {pages} {759} (\bibinfo {year} {2005})}\BibitemShut {NoStop}%
\bibitem [{\citenamefont {M\'ezard}\ \emph
  {et~al.}(2005{\natexlab{a}})\citenamefont {M\'ezard}, \citenamefont
  {Palassini},\ and\ \citenamefont {Rivoire}}]{Mezard-PRL2005}%
  \BibitemOpen
  \bibfield  {author} {\bibinfo {author} {\bibfnamefont {M.}~\bibnamefont
  {M\'ezard}}, \bibinfo {author} {\bibfnamefont {M.}~\bibnamefont {Palassini}},
  \ and\ \bibinfo {author} {\bibfnamefont {O.}~\bibnamefont {Rivoire}},\
  }\bibfield  {title} {\enquote {\bibinfo {title} {Landscape of solutions in
  constraint satisfaction problems},}\ }\href@noop {} {\bibfield  {journal}
  {\bibinfo  {journal} {Phys. Rev. Lett.}\ }\textbf {\bibinfo {volume} {95}},\
  \bibinfo {pages} {200202} (\bibinfo {year} {2005}{\natexlab{a}})}\BibitemShut
  {NoStop}%
\bibitem [{\citenamefont {M\'ezard}\ \emph
  {et~al.}(2005{\natexlab{b}})\citenamefont {M\'ezard}, \citenamefont {Mora},\
  and\ \citenamefont {Zecchina}}]{Mezard-Zecchina-PRL2005}%
  \BibitemOpen
  \bibfield  {author} {\bibinfo {author} {\bibfnamefont {M.}~\bibnamefont
  {M\'ezard}}, \bibinfo {author} {\bibfnamefont {T.}~\bibnamefont {Mora}}, \
  and\ \bibinfo {author} {\bibfnamefont {R.}~\bibnamefont {Zecchina}},\
  }\bibfield  {title} {\enquote {\bibinfo {title} {Clustering of solutions in
  the random satisfiability problem},}\ }\href@noop {} {\bibfield  {journal}
  {\bibinfo  {journal} {Phys. Rev. Lett.}\ }\textbf {\bibinfo {volume} {94}},\
  \bibinfo {pages} {197205} (\bibinfo {year} {2005}{\natexlab{b}})}\BibitemShut
  {NoStop}%
\bibitem [{\citenamefont {Krzakala}\ \emph {et~al.}(2007)\citenamefont
  {Krzakala}, \citenamefont {Montanari}, \citenamefont {Ricci-Tersenghi},
  \citenamefont {Semerjian},\ and\ \citenamefont
  {Zdeborov\'a}}]{Krzakala-PNAS}%
  \BibitemOpen
  \bibfield  {author} {\bibinfo {author} {\bibfnamefont {F.}~\bibnamefont
  {Krzakala}}, \bibinfo {author} {\bibfnamefont {A.}~\bibnamefont {Montanari}},
  \bibinfo {author} {\bibfnamefont {F.}~\bibnamefont {Ricci-Tersenghi}},
  \bibinfo {author} {\bibfnamefont {G.}~\bibnamefont {Semerjian}}, \ and\
  \bibinfo {author} {\bibfnamefont {L.}~\bibnamefont {Zdeborov\'a}},\
  }\bibfield  {title} {\enquote {\bibinfo {title} {Gibbs states and the set of
  solutions of random constraint satisfaction problems},}\ }\href@noop {}
  {\bibfield  {journal} {\bibinfo  {journal} {PNAS}\ }\textbf {\bibinfo
  {volume} {104}},\ \bibinfo {pages} {10318} (\bibinfo {year}
  {2007})}\BibitemShut {NoStop}%
\bibitem [{\citenamefont {Achlioptas}(2008)}]{Achlioptas-EPJB}%
  \BibitemOpen
  \bibfield  {author} {\bibinfo {author} {\bibfnamefont {D.}~\bibnamefont
  {Achlioptas}},\ }\bibfield  {title} {\enquote {\bibinfo {title} {Solution
  clustering in random satisfiability},}\ }\href@noop {} {\bibfield  {journal}
  {\bibinfo  {journal} {Eur. Phys. J. B.}\ }\textbf {\bibinfo {volume} {64}},\
  \bibinfo {pages} {395} (\bibinfo {year} {2008})}\BibitemShut {NoStop}%
\bibitem [{\citenamefont {Ardelius}\ and\ \citenamefont
  {Zdeborova}(2008)}]{PRE_AZ08}%
  \BibitemOpen
  \bibfield  {author} {\bibinfo {author} {\bibfnamefont {J.}~\bibnamefont
  {Ardelius}}\ and\ \bibinfo {author} {\bibfnamefont {L.}~\bibnamefont
  {Zdeborova}},\ }\bibfield  {title} {\enquote {\bibinfo {title} {Exhaustive
  enumeration unveils clustering and freezing in the random 3-satisfiability
  problem},}\ }\href@noop {} {\bibfield  {journal} {\bibinfo  {journal} {Phys.
  Rev. E}\ }\textbf {\bibinfo {volume} {78}},\ \bibinfo {pages} {040101}
  (\bibinfo {year} {2008})}\BibitemShut {NoStop}%
\bibitem [{\citenamefont {Zdeborov\'a}\ and\ \citenamefont
  {M\'{e}zard}(2008)}]{JStatMech_ZM08}%
  \BibitemOpen
  \bibfield  {author} {\bibinfo {author} {\bibfnamefont {L.}~\bibnamefont
  {Zdeborov\'a}}\ and\ \bibinfo {author} {\bibfnamefont {M.}~\bibnamefont
  {M\'{e}zard}},\ }\bibfield  {title} {\enquote {\bibinfo {title} {Constraint
  satisfaction problems with isolated solutions are hard},}\ }\href@noop {}
  {\bibfield  {journal} {\bibinfo  {journal} {J. Stat. Mech.: Theor. Exp.}\ ,\
  \bibinfo {pages} {P12004}} (\bibinfo {year} {2008})}\BibitemShut {NoStop}%
\bibitem [{\citenamefont {Marino}\ \emph {et~al.}(2015)\citenamefont {Marino},
  \citenamefont {Parisi},\ and\ \citenamefont
  {Ricci-Tersenghi}}]{marino2015backtracking}%
  \BibitemOpen
  \bibfield  {author} {\bibinfo {author} {\bibfnamefont {R.}~\bibnamefont
  {Marino}}, \bibinfo {author} {\bibfnamefont {G.}~\bibnamefont {Parisi}}, \
  and\ \bibinfo {author} {\bibfnamefont {F.}~\bibnamefont {Ricci-Tersenghi}},\
  }\bibfield  {title} {\enquote {\bibinfo {title} {The backtracking survey
  propagation algorithm for solving random k-sat problems},}\ }\href@noop {}
  {\bibfield  {journal} {\bibinfo  {journal} {arXiv preprint arXiv:1508.05117}\
  } (\bibinfo {year} {2015})}\BibitemShut {NoStop}%
\bibitem [{url()}]{url_trun_norm_dist}%
  \BibitemOpen
  \href@noop {} {\enquote {\bibinfo {title} {Truncated normal distribution},}\
  }\bibinfo {howpublished}
  {\url{http://en.wikipedia.org/wiki/Truncated_normal_distribution}},\ \bibinfo
  {note} {accessed on Jun 3, 2015}\BibitemShut {NoStop}%
\bibitem [{\citenamefont {Antal}\ \emph {et~al.}(2004)\citenamefont {Antal},
  \citenamefont {Droz},\ and\ \citenamefont {R{\'a}cz}}]{antal2004}%
  \BibitemOpen
  \bibfield  {author} {\bibinfo {author} {\bibfnamefont {T.}~\bibnamefont
  {Antal}}, \bibinfo {author} {\bibfnamefont {M.}~\bibnamefont {Droz}}, \ and\
  \bibinfo {author} {\bibfnamefont {Z.}~\bibnamefont {R{\'a}cz}},\ }\bibfield
  {title} {\enquote {\bibinfo {title} {Probability distribution of
  magnetization in the one-dimensional \uppercase{I}sing model: Effects of
  boundary conditions},}\ }\href@noop {} {\bibfield  {journal} {\bibinfo
  {journal} {J. Phys. A: Math. Gen.}\ }\textbf {\bibinfo {volume} {37}},\
  \bibinfo {pages} {1465} (\bibinfo {year} {2004})}\BibitemShut {NoStop}%
\bibitem [{\citenamefont {Portelli}\ \emph {et~al.}(2001)\citenamefont
  {Portelli}, \citenamefont {Holdsworth}, \citenamefont {Sellitto},\ and\
  \citenamefont {S.T.}}]{PhysRevE_PortelliHSB_01}%
  \BibitemOpen
  \bibfield  {author} {\bibinfo {author} {\bibfnamefont {B.}~\bibnamefont
  {Portelli}}, \bibinfo {author} {\bibfnamefont {P.C.W.}\ \bibnamefont
  {Holdsworth}}, \bibinfo {author} {\bibfnamefont {M.}~\bibnamefont
  {Sellitto}}, \ and\ \bibinfo {author} {\bibfnamefont {Bramwell}\ \bibnamefont
  {S.T.}},\ }\bibfield  {title} {\enquote {\bibinfo {title} {Universal magnetic
  fluctuations with a field-induced length scale},}\ }\href@noop {} {\bibfield
  {journal} {\bibinfo  {journal} {Physical Review E}\ }\textbf {\bibinfo
  {volume} {64}},\ \bibinfo {pages} {036111} (\bibinfo {year}
  {2001})}\BibitemShut {NoStop}%
\bibitem [{\citenamefont {Wang}(2004)}]{thesis_wang}%
  \BibitemOpen
  \bibfield  {author} {\bibinfo {author} {\bibfnamefont {X.}~\bibnamefont
  {Wang}},\ }\emph {\bibinfo {title} {Analysis on the assignment landscape of
  3-\uppercase{SAT} problems}},\ \href
  {http://www.cs.rice.edu/CS/Verification/Theses/Archive/xiaoxumsthesis.pdf}
  {\bibinfo {type} {\uppercase{MS}c \uppercase{T}hesis}},\ \bibinfo  {school}
  {Rice University}, \bibinfo {address} {Houston, TX} (\bibinfo {year}
  {2004})\BibitemShut {NoStop}%
\bibitem [{\citenamefont {Montanari}\ \emph {et~al.}(2008)\citenamefont
  {Montanari}, \citenamefont {Ricci-Tersenghi},\ and\ \citenamefont
  {Semerjian}}]{StatMech_MRTS08}%
  \BibitemOpen
  \bibfield  {author} {\bibinfo {author} {\bibfnamefont {A.}~\bibnamefont
  {Montanari}}, \bibinfo {author} {\bibfnamefont {F.}~\bibnamefont
  {Ricci-Tersenghi}}, \ and\ \bibinfo {author} {\bibfnamefont {G.}~\bibnamefont
  {Semerjian}},\ }\bibfield  {title} {\enquote {\bibinfo {title} {Clusters of
  solutions and replica symmetry breaking in random k-satisfiability},}\
  }\href@noop {} {\bibfield  {journal} {\bibinfo  {journal} {Journal of
  Statistical Mechanics: Theory and Experiment}\ }\textbf {\bibinfo {volume}
  {2008}},\ \bibinfo {pages} {P04004} (\bibinfo {year} {2008})}\BibitemShut
  {NoStop}%
\bibitem [{\citenamefont {M{\'e}zard}\ \emph {et~al.}(2003)\citenamefont
  {M{\'e}zard}, \citenamefont {Ricci-Tersenghi},\ and\ \citenamefont
  {Zecchina}}]{mezard2003two}%
  \BibitemOpen
  \bibfield  {author} {\bibinfo {author} {\bibfnamefont {M.}~\bibnamefont
  {M{\'e}zard}}, \bibinfo {author} {\bibfnamefont {F.}~\bibnamefont
  {Ricci-Tersenghi}}, \ and\ \bibinfo {author} {\bibfnamefont {R.}~\bibnamefont
  {Zecchina}},\ }\bibfield  {title} {\enquote {\bibinfo {title} {Two solutions
  to diluted p-spin models and xorsat problems},}\ }\href@noop {} {\bibfield
  {journal} {\bibinfo  {journal} {Journal of Statistical Physics}\ }\textbf
  {\bibinfo {volume} {111}},\ \bibinfo {pages} {505} (\bibinfo {year}
  {2003})}\BibitemShut {NoStop}%
\bibitem [{\citenamefont {Ott}(2002)}]{ott2002chaos}%
  \BibitemOpen
  \bibfield  {author} {\bibinfo {author} {\bibfnamefont {E.}~\bibnamefont
  {Ott}},\ }\href@noop {} {\emph {\bibinfo {title} {Chaos in dynamical
  systems}}}\ (\bibinfo  {publisher} {Cambridge University Press},\ \bibinfo
  {year} {2002})\BibitemShut {NoStop}%
\bibitem [{\citenamefont {Chv{\'a}tal}\ and\ \citenamefont
  {Reed}(1992)}]{chvatal1992mick}%
  \BibitemOpen
  \bibfield  {author} {\bibinfo {author} {\bibfnamefont {V.}~\bibnamefont
  {Chv{\'a}tal}}\ and\ \bibinfo {author} {\bibfnamefont {B.}~\bibnamefont
  {Reed}},\ }\bibfield  {title} {\enquote {\bibinfo {title} {Mick gets some
  (the odds are on his side)[satisfiability]},}\ }in\ \href@noop {} {\emph
  {\bibinfo {booktitle} {Foundations of Computer Science, 1992. Proceedings.,
  33rd Annual Symposium on}}}\ (\bibinfo {organization} {IEEE},\ \bibinfo
  {year} {1992})\ pp.\ \bibinfo {pages} {620--627}\BibitemShut {NoStop}%
\bibitem [{\citenamefont {Goerdt}(1992)}]{goerdt1992threshold}%
  \BibitemOpen
  \bibfield  {author} {\bibinfo {author} {\bibfnamefont {A.}~\bibnamefont
  {Goerdt}},\ }\bibfield  {title} {\enquote {\bibinfo {title} {A threshold for
  unsatisfiability},}\ }in\ \href@noop {} {\emph {\bibinfo {booktitle}
  {Mathematical foundations of computer science 1992}}}\ (\bibinfo  {publisher}
  {Springer},\ \bibinfo {year} {1992})\ pp.\ \bibinfo {pages}
  {264--274}\BibitemShut {NoStop}%
\bibitem [{\citenamefont {Beeri}\ and\ \citenamefont
  {Bernstein}(1979)}]{Beeri_1979}%
  \BibitemOpen
  \bibfield  {author} {\bibinfo {author} {\bibfnamefont {C.}~\bibnamefont
  {Beeri}}\ and\ \bibinfo {author} {\bibfnamefont {P.A.}\ \bibnamefont
  {Bernstein}},\ }\bibfield  {title} {\enquote {\bibinfo {title} {Computational
  problems related to the design of normal form relational schemas},}\
  }\href@noop {} {\bibfield  {journal} {\bibinfo  {journal} {ACM Trans.
  Database Syst.}\ }\textbf {\bibinfo {volume} {4}},\ \bibinfo {pages} {30}
  (\bibinfo {year} {1979})}\BibitemShut {NoStop}%
\bibitem [{\citenamefont {Dowling}\ and\ \citenamefont
  {Gallier}(1984)}]{dowling1984linear}%
  \BibitemOpen
  \bibfield  {author} {\bibinfo {author} {\bibfnamefont {W.F}\ \bibnamefont
  {Dowling}}\ and\ \bibinfo {author} {\bibfnamefont {J.H.}\ \bibnamefont
  {Gallier}},\ }\bibfield  {title} {\enquote {\bibinfo {title} {Linear-time
  algorithms for testing the satisfiability of propositional \uppercase{H}orn
  formulae},}\ }\href@noop {} {\bibfield  {journal} {\bibinfo  {journal} {The
  Journal of Logic Programming}\ }\textbf {\bibinfo {volume} {1}},\ \bibinfo
  {pages} {267} (\bibinfo {year} {1984})}\BibitemShut {NoStop}%
\bibitem [{\citenamefont {Moore}\ \emph {et~al.}(2005)\citenamefont {Moore},
  \citenamefont {Istrate}, \citenamefont {Demopoulos},\ and\ \citenamefont
  {Vardi}}]{moore2005continuous}%
  \BibitemOpen
  \bibfield  {author} {\bibinfo {author} {\bibfnamefont {C.}~\bibnamefont
  {Moore}}, \bibinfo {author} {\bibfnamefont {G.}~\bibnamefont {Istrate}},
  \bibinfo {author} {\bibfnamefont {D.}~\bibnamefont {Demopoulos}}, \ and\
  \bibinfo {author} {\bibfnamefont {M.Y.}\ \bibnamefont {Vardi}},\ }\bibfield
  {title} {\enquote {\bibinfo {title} {A continuous-discontinuous second-order
  transition in the satisfiability of random \uppercase{H}orn-\uppercase{SAT}
  formulas},}\ }in\ \href@noop {} {\emph {\bibinfo {booktitle} {Approximation,
  Randomization and Combinatorial Optimization. Algorithms and Techniques}}}\
  (\bibinfo  {publisher} {Springer},\ \bibinfo {year} {2005})\ pp.\ \bibinfo
  {pages} {414--425}\BibitemShut {NoStop}%
\end{thebibliography}
%

\end{document}